\documentclass[aps,prd,reprint,a4paper,showpacs,nofootinbib,superscriptaddress]{revtex4-2}

\usepackage{bbm}
\usepackage{amsmath}
\usepackage{graphicx}
\usepackage{float}
\usepackage{amssymb}
\usepackage{subfigure}
\usepackage{amssymb}
\usepackage{hyperref}
\usepackage{epstopdf}

\usepackage[utf8]{inputenc}
\hypersetup{
    colorlinks=true,
    linkcolor=red,
    citecolor=blue,
}
\usepackage{color}
\usepackage[T1]{fontenc}
\usepackage{txfonts}
\usepackage{mathrsfs}
\usepackage{latexsym}
\usepackage{epsfig}
\usepackage{epstopdf}
\usepackage{epstopdf}
\usepackage{graphicx}
\usepackage{amssymb}
\usepackage{amsmath}
\usepackage{dcolumn}
\usepackage{bm}
\usepackage{color}
\usepackage{comment}
\usepackage{xcolor}
\usepackage{dsfont}

\begin{document}
\title{\bf Analytical and numerical study of accretion processes around charged spherically symmetric black holes in scalar-tensor Gauss–Bonnet gravity}

\author{G. Mustafa}
\email{gmustafa3828@gmail.com }
\affiliation{Department of Physics, Zhejiang Normal University,
Jinhua 321004, China}

\author{Orhan~Donmez}
\email{orhan.donmez@aum.edu.kw}
\affiliation{College of Engineering and Technology, American University of the Middle East, Egaila 54200, Kuwait}

\author{A. Errehymy}
\email{abdelghani.errehymy@gmail.com}\affiliation{Astrophysics Research Centre, School of Mathematics, Statistics and Computer Science, University of KwaZulu-Natal, Private Bag X54001, Durban 4000, South Africa}

\author{Faisal Javed}
\email{faisaljaved.math@gmail.com}\affiliation{Department of
Physics, Zhejiang Normal University, Jinhua 321004, People's
Republic of China}

\author{A.~Ditta}
\email{adsmeerkhan@gmail.com}
\affiliation{Research Center of Astrophysics and Cosmology, Khazar University, Baku, AZ1096, 41 Mehseti Street, Azerbaijan}

\author{Tayyab Naseer}
\email{tayyabnaseer48@yahoo.com}\affiliation{Department of
Mathematics and Statistics, The University of Lahore,\\
1-KM Defence Road Lahore-54000, Pakistan}

\author{S.K. Maurya}
\email{sunil@unizwa.edu.om}\affiliation{Department of Mathematical and Physical Sciences, College of Arts and Sciences, University of Nizwa, Nizwa 616, Sultanate of Oman}

\author{Farruh Atamurotov}
\email{atamurotov@yahoo.com}
\affiliation{Urgench State University, Urgench 220100, Uzbekistan}

\begin{abstract}
We investigate the physical phenomena occurring around a spherically symmetric, non-rotating charged black hole (BH) to explore the effects of scalar–tensor Gauss–Bonnet gravity on circular motion, accretion disk properties, and Bondi–Hoyle–Lyttleton (BHL) accretion flow. By analytically and numerically examining the influence of the Gauss–Bonnet coupling constant $c_1$ and the cosmological parameter $\Lambda$, we reveal how these modified gravity parameters alter the underlying physical processes. Using geodesic analysis, we compute the specific energy, angular momentum, innermost stable circular orbit (ISCO) radius, and radiation flux of test particles, providing insight into how the modified gravity framework affects orbital stability and the organization of the accretion flow. Subsequently, through numerical solutions of the general relativistic hydrodynamic (GRHD) equations, we describe the morphology of the shock cone formed via the BHL accretion mechanism around the BH. The numerical results demonstrate that increasing the values of $c_1$ and negative $\Lambda$ reduce gravitational focusing. Consequently, depending on the parameter choices, the opening angle of the shock cone either widens or narrows compared to the Schwarzschild case. However, because of weakened gravitational focusing, both the amount of accreted matter and the density of material trapped inside the cone decrease significantly. These results indicate that scalar–tensor Gauss–Bonnet corrections act as an effective gravitational damping term, transferring turbulence and transforming shock-dominated accretion into more stable configurations. The consistency between theoretical and numerical results suggests that the observable properties of accretion disks and quasi-periodic oscillations (QPOs) can serve as probes to constrain the parameters of scalar–tensor Gauss–Bonnet gravity in strong-field regimes.

\end{abstract}

\maketitle

\date{\today}

\section{Introduction}

The special properties of BH are a consequence of general relativity (GR) and recent advances in the understanding of their astrophysical implications. Specifically, BHs are distinguished by extremely strong spins, magnetic fields, and gravitational pulls. As a result, they provide unique opportunities for the investigation of the relationship between gravity and matter. The first evidence for the existence of BHs was the gravitational waves (GW) detected from the collision of a pair of binary BHs, a finding made possible by the joint work of the LIGO and Virgo gravitational-wave observatories \cite{1l}. The Event Horizon Telescope (EHT) made another unprecedented achievement in astrophysics by employing very long baseline interferometry to obtain the first photograph of a BH shadow located in the $M87^{\ast}$ galaxy. Additionally, a notable milestone was reached with the release of an image depicting the BH Sgr $A^{\ast}$ at the center of the Milky Way \cite{2l,3l}. The electromagnetic spectrum emitted by accretion disks has provided additional evidence for the existence of BHs \cite{5,6,7}.  The understanding of GR theory and the formation of accretion disks around supermassive BH has improved due to the new advancements, especially within the strong gravitational field, near the BH’s event horizon. There are also insights for the study of modified theories of gravity. Modified theories of gravity to cosmology and astrophysics which extends beyond GR has garnered considerable interest. In addressing problems such as the accelerated expansion of the universe, the foundations of GR, and the various unsolved problems within these areas, such theories  have also been developed to bridge other gaps in the respective fields \cite{a3,a4,a5,a6,a12}. The exploration of modified gravity theories, which aims to unlock new understanding of the basic nature of gravity and the universe's fundamental structure, has become a prominent area of inquiry \cite{a11}. Nevertheless, the attempts to quantify gravity and unify classical gravity with quantum fields continue to pose a challenge.

Investigating compact objects in the cosmos provides an exceptional opportunity to test extended theories of gravitation and may reveal GR anomalies through strong-gravity effects. In particular, we concentrate on a quantum gravity-inspired theory that incorporates higher-curvature implications in the form of the quadratic Gauss-Bonnet (GB) component \cite{zwiebach1985curvature,gross1987quartic, metsaev1987order}. This theory, known as Einstein-scalar-Gauss-Bonnet (EsGB) theory, in four-dimensional spacetime involves coupling to a scalar field and is distinguished by the selection of the coupling function. Intriguing phenomena including a minimum mass \cite{kanti1998dilatonic}, a violation of the Kerr constraint \cite{kleihaus2011rotating}, and the occurrence of scalarization for different types of coupling function have been discovered as a result of extensive study on EsGB BHs \cite{sotiriou2014black, doneva2018new,blazquez2018radial,collodel2020spinning,berti2021spin}. An exciting result of including the GB component in the theory is the creation of an effective stress-energy tensor that has the potential to violate energy conditions. As a result, wormhole solutions are seen in EsGB theories, which is a notable departure from GR, which requires exotic matter for such solutions \cite{kanti2011wormholes,kanti2012stable,antoniou2020novel}. The intriguing possibility of gravitational particle-like solutions with perfectly regular spacetime is raised by the finding of novel occurrences in EsGB theories, which gives these theories an exceptional feature.

Due to the effect of gravity, the celestial bodies in the universe rotate in patterns and move in orbits, leading to the formation of accretion disks. These systems include very different types of celestial bodies. Systems consist of young stellar bodies that sit in the active galactic nuclei, supermassive BHs, and neutron stars. Although the orbital materials of the accretion disks exhibit stability most of the time, they can become unstable to trigger accretion or mass gain. Generally, the most massive core objects, such as BHs, emanate accretion fluids and gain mass from surrounding fluids. To resolve the problem, one must assess the geodesic motion of particles near BHs, focusing on the ISCO and marginally bound orbits. These distances are instrumental in the study of the BHs accretion disks, as they correspond to the inner edge of the disk \cite{8}. The ongoing advancements in accretion theory are mainly due to research efforts devoted to obtaining accurate, analytical solutions and other contributing factors \cite{bondi1952spherically}. These solutions furnish a precious framework for categorizing various astrophysical scenarios, thereby enhancing our understanding of the phenomena under study. In addition, analytical solutions play a vital role in the comparison and validation of numerical codes, making them indispensable tools in this field \cite{olivares2020tell}. Bondi's model, originating in Newtonian gravity, represents the stationary flow of a spherically symmetric fluid that accretes onto a BH \cite{bondi1952spherically}. Based on this model, researchers have extended it to the relativistic domain by assuming a Schwarzschild BH \cite{michel1972accretion}. This extension offers a more comprehensive understanding of the accretion process in light of GR.

The investigation of BHs has expanded to encompass studies on quantum gravity, dark matter phenomenotechnology, and horizon-scale investigations aside from the classical parameters of general relativity. Some of the interest in reconciling quantum spacetime geometry has been focused on regular BH models that do not admit curvature singularities. Modified BH thermodynamics due to changes in Hawking BH radiation, which has been framed by Garfinkle, among many others, addresses non-minimal couplings of scalars and curvature  \cite{v1}. Moreover, primordial regular BHs originating from effective quantum-gravity theories have been proposed as candidates for black matter, considering both tri-symmetric and non-symmetric spacetime structures \cite{v2,v3,v4}. The study of the dynamics of BHs also includes the gravitational wave signatures that BHs emit, which is particularly prominent in the case of dark energy and cosmological evolution \cite{v5}. These concepts have spurred an investigation of the interaction between BHs and large-scale structures.  The many groundbreaking discoveries from the EHT have opened up new opportunities for testing fundamental theories of physics using the BH shadows. Horizon-scale imaging has been used to rule out BH solutions in mimetic gravity \cite{v6}, carry out the most detailed constraints on the beyond-GR effects around Sgr A, and investigate the possible presence of ultralight bosonic particles via shadow distortions and superradiant instabilities. In addition, studies on rotating regular BHs have uncovered significant connections between quasi-normal modes and shadow features \cite{v7}. Many researchers investigated the shadows of BHs with scalar hair \cite{v8}, their role as standard candles in cosmology \cite{v9}, and the shadow resulting from the presence of extra dimensions \cite{v10} or possible overspinning states of M87 \cite{v11}.

In the relativistic context of a Schwarzschild BH, researchers have made significant contributions by establishing analytical solutions for wind accretion scenarios \cite{bondi1944mechanism,hoyle1939effect}. The understanding of spherical and wind accretion has been significantly advanced through a multitude of investigations, covering both analytical and numerical approaches \cite{karkowski2013bondi, chaverra2015radial}. Among the researchers in this field, Sharif and his colleagues have investigated phantom accretion in a particular class of BHs, making notable contributions to the understanding of this intriguing phenomenon \cite{sharif2011phantom,sharif2012phantom}.  Ditta and Abbas have also offered valuable insights into accretion processes, considering various types of BH \cite{ditta2020circular,ditta2020relativistic,ditta2020astrophysical,feng2024analysis,caliskan2024particle}. Investigations have been extended to modified theories of gravity, including quantum gravity corrections, non-commutative theory, and accretion flows on scalar-tensor-vector gravity \cite{john2019black, gangopadhyay2018accretion, yang2015quantum}. In addition, the study of cyclic and heteroclinic accretion processes has been conducted on $f(R)$ and $f(T)$ BH \cite{ahmed2016astrophysical}. In addition, researchers have explored the study of accretion fluxes in static BHs using dynamical systems, as evidenced by the work listed in \cite{jawad2017accreting,abbas2018accretion,abbas2019matter,abbas2019mattera,abbas2020matter}.

The current study focuses on exploring the geodesic behavior of the charged BH in the framework of the scalar tensor EsGB. We also study the accretion disk of the BH geometry considered. An outline of the structure may be seen below: We present a comprehensive review of charged BH in the scalar tensor EsGB in Section II. Section III is devoted to study the geodesic behavior, circular motion, radiation energy flux, and oscillations. Section IV deals with the accretion process for considering the BH geometry. Section V is used to explore the epicyclic frequencies of the considered BH structure. Section IV presents a detailed analysis related to the dynamics of BHL accretion in EsGB charged BH scalar tensor. In the last section, we present the concluding remarks of our presented analysis.

\section{charged black hole in scalar-tensor Gauss-Bonnet gravity}

This section mainly deals with the framework related to the considered black hole solution. After adopting the procedure previously reported \cite{Capozziello:2023vvr}, we have the following Lagrangian:

\begin{eqnarray}\label{ll1}
S &&=\int \mathrm{d}^4 x \sqrt{-g}(\frac{1}{2 \kappa^2} R+\lambda(\frac{1}{2} \partial_\mu \zeta \partial^a \zeta+\frac{a^4}{2})\\&&+h(\zeta) \mathcal{G}-\tilde{V}(\zeta)+\mathcal{L}_{\text {matter }}).\nonumber
\end{eqnarray}

The Lagrangian given in the above equation can get the following two equations of motion:
\begin{widetext}
\begin{eqnarray}
& 0=-\frac{1}{\sqrt{-g}} \partial_a\left(\lambda \omega(\zeta) g^{a b} \sqrt{-g} \partial_b \zeta\right)+h^{\prime}(\zeta) \mathcal{G}-\tilde{V}^{\prime}(\zeta)+\frac{1}{2} \lambda \omega^{\prime}(\zeta) g^{a b} \partial_a \zeta \partial_b \zeta \\
& 0=\frac{1}{2 \kappa^2}\left(-R_{a b}+\frac{1}{2} g_{a b} R\right)+\frac{1}{2} T_{\text {matter } a b}-\frac{1}{2} \lambda \partial_a \zeta \partial_b \zeta-\frac{1}{2} g_{a b} \tilde{V}(\zeta)+D_{a b}^{\tau \eta} \nabla_\tau \nabla_\eta h(\zeta)
\end{eqnarray}
\end{widetext}
Let us rewrite the action from Eq. (\ref{ll1}) by including electromagnetic field as:

\begin{eqnarray}\label{ll2}
\mathcal{L}&&=\int \mathrm{d}^4 x \sqrt{-g}\left[\frac{1}{2 \kappa^2} R+\lambda\left(\frac{1}{2} \omega(\zeta) \partial_a \zeta \partial^a \zeta+\frac{a^4}{2}\right)\right.\\&&+h(\zeta) \mathcal{G}-\tilde{V}(\zeta)+\mathcal{L}_{\text {matter }}-\Lambda+\mathcal{L}_{\mathrm{em}}\left.\right] .\nonumber
\end{eqnarray}

\begin{figure*}
\centering \epsfig{file=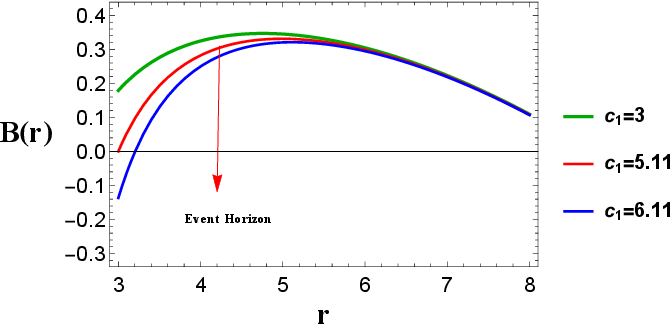, width=.45\linewidth,
height=2.in}\;\;\;\;\;\;\;\;\epsfig{file=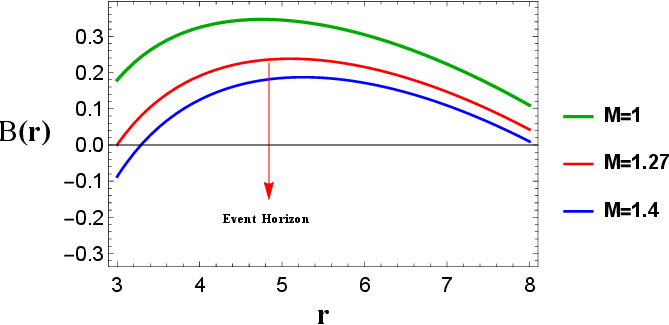, width=.45\linewidth,
height=2.in}
\caption{\label{F2} The graph displays the horizons of the CBH, represented by $B(r)$, according to the radial coordinate $r$. It shows the variation in horizons for different values of the parameters $M$, $c_1$ and $\Lambda$.}
\end{figure*}

The Lagrangian for the electromagnetic field, $\mathcal{L}_{\mathrm{em}}$, is expressed as:
\begin{equation}\label{ll3}
\mathcal{L}_{\mathrm{em}}=-\frac{1}{2} \mathcal{F}\left(\mathcal{F} \equiv \mathcal{F}_{a b} \mathcal{F}^{a b}\right).
\end{equation}
Now, after some calculations, which are already reported in \cite{Capozziello:2023vvr}, we have the following charged black hole solution in scalar-tensor Gauss-Bonnet gravity:
\begin{equation}
ds^{2}=-A(r)dt^{2}+\frac{1}{B(r)}dr^{2}+r^{2}
(d\theta^{2}+ \sin^{2}\theta d\phi^{2}),\label{1}
\end{equation}
where
\begin{eqnarray}\label{2}
A(r)&=& \Lambda r^2+1-\frac{2M}{r}+\frac{2Mc_1}{r^3}-\frac{2Mc^2_1}{r^5},\\
\label{3}
B(r)&=& \Lambda r^2+1-\frac{2M}{r}+\frac{c^{5/2}_1}{r^5}.
\end{eqnarray}
To formulate the horizon radii of the solution (\ref{1}), we need to consider $B(r)=0$ which gives seven roots. Analyzing the exact analytic expressions for these solutions can be challenging. However, we can deduce the asymptotic expressions and plot them using some suitable numerical values of the parameters $c_1$ and $\Lambda$ that characterize the model. The graph depicted in Fig. (\ref{F2}) shows the effect of various parameters on the metric function $B(r)$, in particular on the horizon. The graph highlights several key points, which are as follows: In the upper left plot of Fig. (\ref{F2}), we can see that among the curves representing different values of $c_1$, only the red curve shows an event horizon. This happens specifically at $c_1 = 5.11$, while the values of $M=1$ and $\Lambda=-0.01$ are kept fixed. From the upper right plot in Fig. (\ref{F2}), it is clear that only the red curve shows an event horizon. This is observed when the parameter $M$ takes the value of $1.27$ while holding other parameters such as $c_1$ at $3$ and $\Lambda$ at $-0.01$.

In the bottom plot of Fig. (\ref{F2}), we notice the presence of a singularity where the metric function $B(r)$ goes to zero. However, the metric potential $A(r)$ remains positive and is very far from the singularity. This means that the singularity is located far from the zone where $A(r)$ remains sharply defined and positive. Upon examining the horizon behavior of the CBH horizon, it is obvious that the curves display an inward shift towards an event horizon with increasing values of $M$ and $c_1$. Consequently, whenever $M$ and $c_1$ become larger, the event horizon of the CBH is moved to a smaller radial distance.

\section{geodesic motion around Charged  black holes in scalar-tensor Gauss-Bonnet gravity} 

The general solutions for the geodesic motion of particles in static and spherically symmetric spacetime are determined by considering the two Killing vectors $\varepsilon_t=\partial_t$ and $\varepsilon_\phi=\partial_\phi$. These vectors are associated with the conserved quantities of energy ($E$) and angular momentum ($L$) along the particle's trajectory. Hence, the path of the particle can be described by the following relations
\begin{eqnarray}
E&=&-g_{\mu\nu}\varepsilon^\mu _t u^\nu\equiv-u_t,\\
\label{9}
L&=&g_{\mu\nu}\varepsilon^\mu _\phi u^\nu\equiv u_\phi.
\end{eqnarray}
Taking into account the $4$-velocity as $u^\mu = \frac{dx^\mu}{d\tau}=(u^t,u^r,u^\theta,u^\phi)$, where $\tau$ is the proper time, we can impose the normalization condition. This condition yields the following constraint
\begin{equation}
g_{rr}(u^r)^2+g^{tt}(u_t)^2=-\Big[1-g_{\theta\theta}(u^\theta)^2-g^{\phi\phi}(u_\phi)^2\Big].\label{10}
\end{equation}
By considering the equatorial plane ($\theta=\frac{\pi}{2}$) and using Eqs. (\ref{9}) and (\ref{10}), we can extract the following expressions
\begin{eqnarray}\label{11}
u^t&=&\frac{E}{A(r)},\\
u^\theta&=&0,\\
u^\phi&=&\frac{L}{r^2},\\
u^r&=&\sqrt{B(r)\left(-1+\frac{E^2}{A(r)}-\frac{L^2}{r^2}\right)}.
\end{eqnarray}
By using Eq. (\ref{11}), we obtain the formula for the conserved energy of a particle's motion, which incorporates an effective potential $(V_{eff})$. This formula can be expressed as follows
\begin{equation}
E^2=\frac{A(r)}{B(r)}(u^r)^2+V_{eff},\label{12}
\end{equation}
with the effective potential given by
\begin{equation}
V_{eff}=A(r)\left[1+\frac{L^2}{r^2}\right].\label{13}
\end{equation}

\begin{figure*}
\centering \epsfig{file=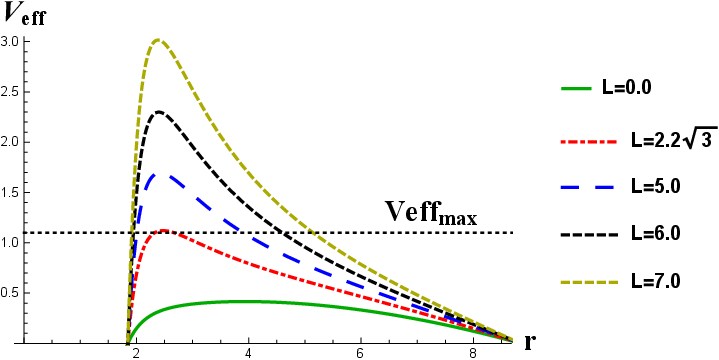, width=.40\linewidth,
height=2.0in}~~~~~~~~~~~~\epsfig{file=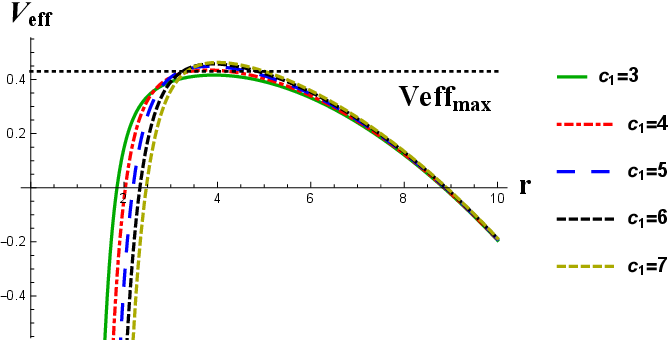, width=.40\linewidth,
height=2.0in}
\centering \epsfig{file=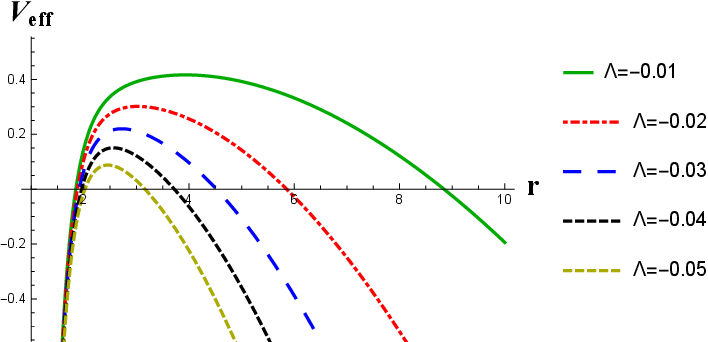, width=.40\linewidth,
height=2.0in}\caption{\label{F3} The graph shows the variation of the effective potential $V_{eff}$ of the CBH according to $r$, considering different values of $c_1$ and $\Lambda$.}
\end{figure*}

The features of the effective potential for the CBH play a pivotal role in explaining geodesic motion, as they are affected by the parameter $A(r)$ and the angular momentum. To localize the circular orbits, we investigate the local extremums of this effective quantity. These extremums specify the points at which stable circular orbits can exist. Figure (\ref{F3}) highlights some important points In the plot on the left, we can see that the dark green curve does not exhibit any extrema for $L=0$. The first extremum is detected in the red curve at $V_{eff}=1.1$ for $L=2.2\sqrt{3}$. Clearly, in this plot, $V_{eff}$ increases with $L$. The values of the other parameters are taken as $M=1$, $\Lambda=-0.01$, and $c_1=3$.

In the right plot, we observe that the first extremum is evident for $c_1=3$. As $c_1$ increases, $V_{eff}$ decreases, and the curves diverge further from the source. The values of the other parameters used in this graph are $M=1$, $L=5$, and $\Lambda=-0.01$. In the bottom plot, we can observe that all curves exhibit a decreasing trend and approach the source as $\Lambda$ increases. As this parameter increases, the effective potential decreases as well. The values of the other parameters utilized in this graph are $M=1$, $L=5$, and $c_1=3$.

\subsection{Circular Motion}

In the context of the circular motion of a particle within the context of the equatorial plane, it is assumed that the radial component $r$ remains constant. This suggests that both the radial velocity $u^r$ and its derivative $\dot{u}^r$ are equal to zero, implying that there is no radial motion or a change in the radial velocity. From Eq. (\ref{12}), we can determine that the effective potential is equal to the square of the total energy, i.e., $V_{eff} = E^2$. Additionally, by taking the derivative of $V_{eff}$ with respect to $r$ and setting it to zero, we find $dV_{eff}/dr = 0$. Based on these relationships, we are able to express the specific energy, the specific angular momentum, the angular velocity $\Omega_\phi$ and the angular momentum $l$ as follows
\begin{eqnarray}\label{14}
E^2&=&\frac{2A^{2}(r)}{2A(r)-r A'(r)},\\
\label{15}
L^2&=&\frac{r^2 A'(r)}{2A(r)-rA^{'}(r)},\\
\label{16}
\Omega_{\phi}&=&\frac{d\phi}{dt}\equiv\frac{u^\phi}{u^t}\quad\Rightarrow\quad \Omega^2_{\phi}=\frac{A'(r)}{2r},\\
\label{17}
l^2&=&\frac{L^2}{E^2}=\frac{r^3A'(r)}{2A^2(r)}.
\end{eqnarray}
In order to measure specific energy and angular momentum, we need the following condition 
\begin{equation}
r\Big[2A(r)-rA'(r)\Big]>0.\label{18}
\end{equation}
By solving this condition, one can obtain the limited area of the circular orbit. In addition, $E^2<1$ refers to the bound and $E^2=1$ represents the marginally bound orbits. Using Eq. (\ref{14}), the later kind of orbits are found as
\begin{equation}
r\Big[rA'(r)+2A(r)\left(A(r)-1\right)\Big]=0.\label{19}
\end{equation}
The momentum and energy show divergence at $r$ when using Eqs. (\ref{14}) and (\ref{15}). The relationship for this divergence turns out to be
\begin{equation}
r\Big[2A(r)-r^2A'(r)\Big]=0.\label{20}
\end{equation}
It must be noted that the photon sphere is characterized by this relation.  
\begin{figure*}
\centering \epsfig{file=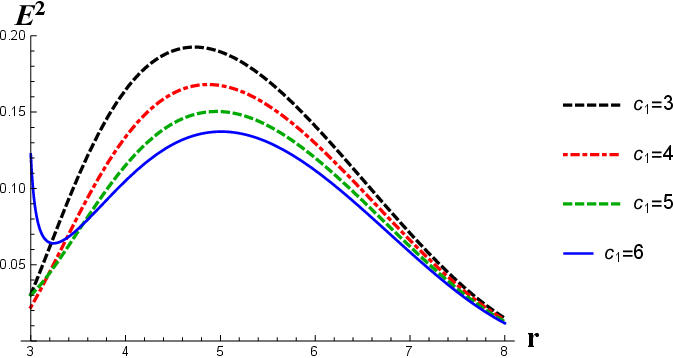, width=.40\linewidth,
height=2.0in}~~~~~~~~~~~~\epsfig{file=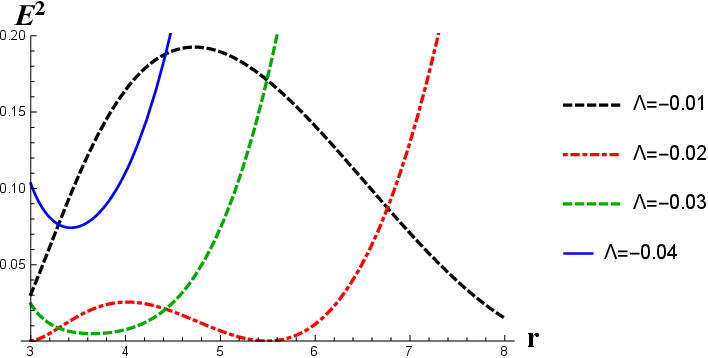, width=.40\linewidth,
height=2.0in}\caption{\label{F4} The graph illustrates the variation of the specific energy $E$ of the CBH with respect to $r$ for different values of $c_1$ and $\Lambda$.}
\end{figure*}
The behavior of a specific energy can be seen in Fig. (\ref{F4}), highlighting several key points. By decreasing the parameter $c_1$, the specific energy increases progressively. This is evident in the left plot of Fig. (\ref{F4}), where the curves are shifted away from the source. As we see from the right plot, the energy decreases as one moves closer to the source, implying  a decrease in energy near the source. In contrast, the energy increases as we move farther away from the source, indicating an increase in energy in these regions. This behavior is directly related to the parameter $\Lambda$.

\begin{figure*}
\centering \epsfig{file=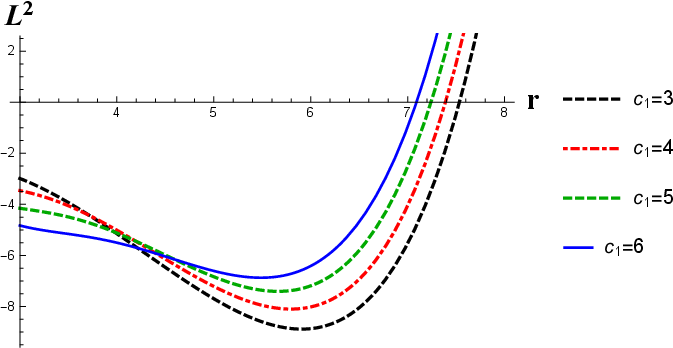, width=.40\linewidth,
height=2.0in}~~~~~~~~~~~~\epsfig{file=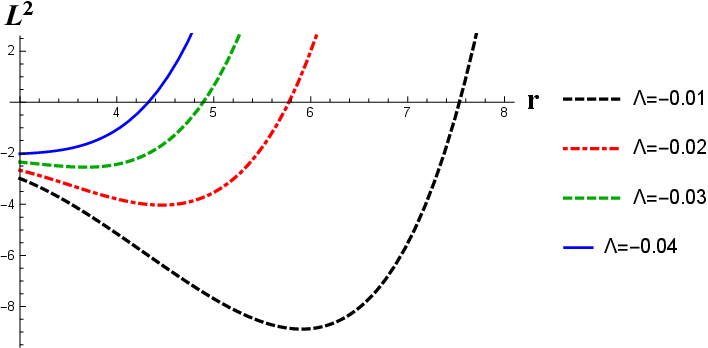, width=.40\linewidth,
height=2.0in}\caption{\label{F5} The graph illustrates the variation of angular momentum $L$ of the CBH with respect to $r$ for different values of $c_1$ and $\Lambda$.}
\end{figure*}

In addition, the angular momentum is characterized in Fig. (\ref{F5}). In the left graph of Fig. (\ref{F5}), it is clear that the angular momentum decreases with decreasing parameter $c_1$. What is more is that the angular momentum decreases with distance from the source, whereas it increases when approaching the BH. The right graph of Fig. (\ref{F5}) shows a striking tendency: as the parameter $\Lambda$ increases, the angular momentum also increases. In particular, as we move closer to the BH, the angular momentum increases significantly, while it decreases as we move farther away from the source.

 \subsection{Radiation Energy Flux}

In accretion analysis, when elements fall from rest to infinity onto the BH, the gravitational energy they release can be converted into radiation. This radiation is responsible for some of the most energetic astrophysical phenomena. The radiation energy flux over the accretion disk is ascribed to the radiation energy combined with the specific energy, angular momentum, and angular velocity, as investigated by Kato et \textit{al.} \cite{10}. Mathematically, we have the following:
\begin{equation}
K=-\frac{\dot{M}\frac{d\Omega_\phi}{dr}}{4\pi \sqrt{-g}(E-L\Omega_\phi)^2}\int(E-L\Omega_\phi)\frac{d}{dr}Ldr.\label{22}
\end{equation}
Here, the radiation flux is represented by $K$ and the accretion rate is denoted as $\dot{M}$. In addition, the parameter $g$ can be expressed as $g=det(g_{\mu\nu})$. Here, it must be recalled that we are confined to the equatorial plane, i.e., $\sin\theta=1$. Using Eqs. (\ref{13})-(\ref{16}), we can obtain the following.
\begin{eqnarray}\label{24}
K(r)=&&-\frac{\dot{M}}{4\pi r^4}\sqrt{\frac{r B(r)}{2A'(r)A(r)}}
 \frac{X(r) W(r)}{Y(r)^2}
\int^{r}_{mb}Z(r)dr,
\end{eqnarray}
where 
\begin{eqnarray}
W(r) &=& \Big[2A(r)-r A'(r)\Big], \\
X(r) &=& \Big[r A''(r)-A'(r)\Big], \\
Y(r) &=& \Big[2A(r)+r A'(r)\Big], \\
Z(r)&=&\sqrt{\frac{r}{2A'(r)}}\frac{Y(r)\Big[-A''(r)r A(r)+2r A'^2(r)-3A'(r)A(r)\Big]}{W(r)^2}.~~\label{25}
\end{eqnarray}
We are also introducing a new relationship $K(r)=\sigma T^4(r)$, which relates the radiation flux to the temperature. Here, $\sigma$ stands for Stefan's constant. The complete analysis of this relationship has been elucidated by Torres and Diego \cite{51} and extended by Torres and Diego Babichev et \textit{al.} \cite{52}.
\begin{figure*}
\centering \epsfig{file=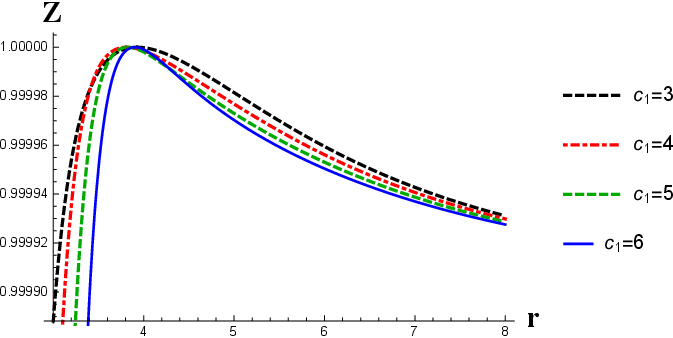, width=.40\linewidth,
height=2.0in}~~~~~~~~~~~~\epsfig{file=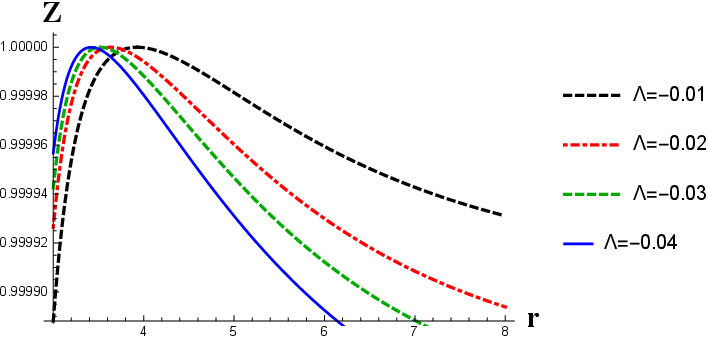, width=.40\linewidth,
height=2.0in}\caption{\label{F6} The graph illustrates the variation of $Z(r)$ of the CBH with respect to $r$ for different values of $c_1$ and $\Lambda$.}
\end{figure*}
In Fig. (\ref{F6}), we can observe the following key points concerning the behavior of $Z$ for the CBH according to $r$ as:  In the left graph, we can see that for decreasing values of $c_1$, all solution curves progressively move away from the source and diverge. In the right graph, as the parameter $\Lambda$ increases, the curves show a remarkable behavior. They rapidly converge at the source, leading to a decrease in radius. Nevertheless, it is interesting to note that the dotted black curve differs from the others in that it does not fall on the source as quickly as the rest of the curves.

\begin{figure*}
\centering \epsfig{file=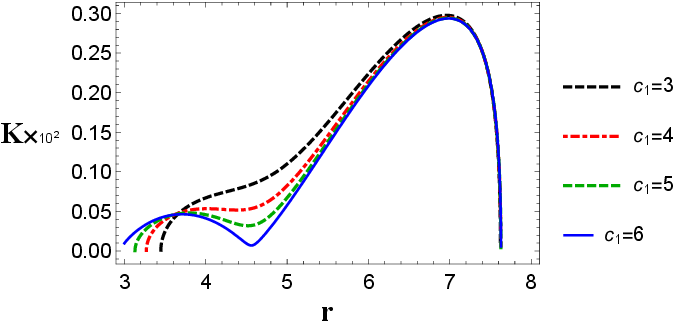, width=.45\linewidth,
height=2.0in}~~~~~~~~~~~~\epsfig{file=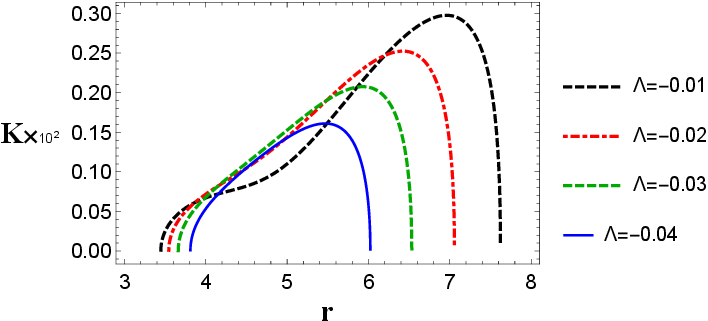, width=.45\linewidth,
height=2.0in}
\centering \epsfig{file=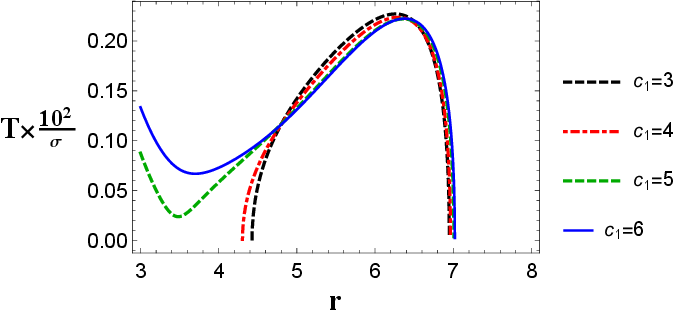, width=.45\linewidth,
height=2.0in}~~~~~~~~~~~~\epsfig{file=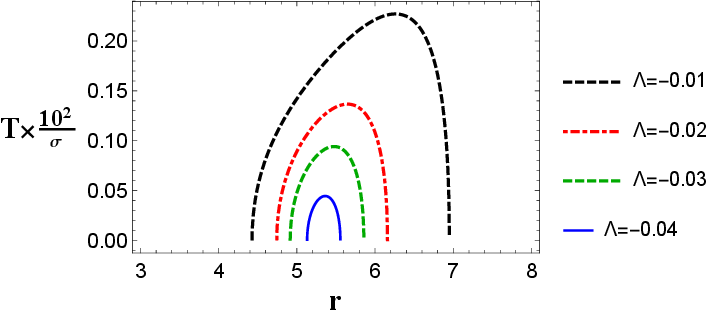, width=.45\linewidth,
height=2.0in}\caption{\label{F7} The graph illustrates the variation of the radiation energy $K(r)$, and the temperature $T$ of the CBH according to $r$ for different values of the parameters $c_1$ and $\Lambda$.}
\end{figure*}

In Fig. (\ref{F7}), both the radiation energy flux and temperature are plotted, providing a comprehensive understanding of the intricate relationship between these fundamental factors. In the upper left plot, all solution curves start from the source and exhibit a decrease in flux as  the parameter $c_1$ increases. In the upper right plot, the decrease in radius near the horizon reduces the flux corresponding to the parameter $\Lambda$, and the curves intersect far from the source.In the lower left plot, the maximum temperature is shown for the initial value of $c_1$, and it decreases with increasing $c_1$. In the lower right plot, the maximum temperature is displayed for the initial value of $\Lambda$ far from the source. It can be seen that the maximum temperature decreases with increasing $\Lambda$.

The analysis of the accretion disk incorporates an invaluable criterion: the efficiency of the accreting particle. To maximize efficiency in the accretion process, the relation $\eta=1-E$ is employed, where $\eta$ stands for efficiency.
\begin{figure*}
\centering \epsfig{file=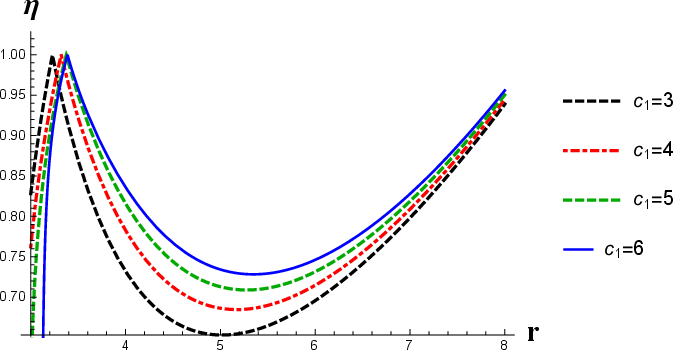, width=.40\linewidth,
height=2.0in}~~~~~~~~~~~~\epsfig{file=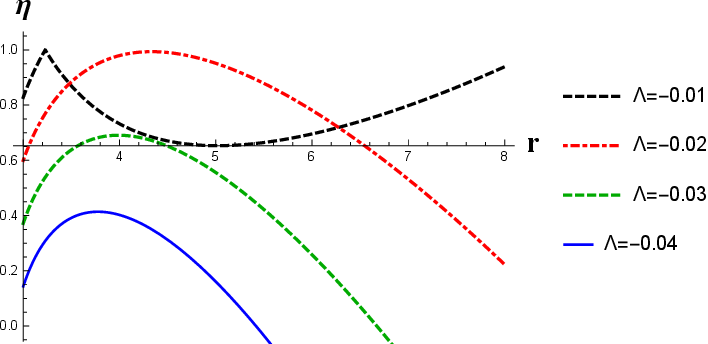, width=.40\linewidth,
height=2.0in}\caption{\label{F8} The graph illustrates how the efficiency of the CBH varies according to $r$, taking into account different values for the parameters $c_1$ and $\Lambda$.}
\end{figure*}
The accreting efficiency corresponding to the metric parameters is shown in Fig. (\ref{F8}). The highlights are 
\begin{itemize}
  \item The plot on the left shows an increasing trend in efficiency with increasing distance from the source mass. However, as the distance approaches the source, the efficiency decreases. These findings are consistent with variations in the parameter $c_1$.
  \item In the right graph, the efficiency gradually decreases as the parameter $\Lambda$ increases. However, at the initial value of $\Lambda$, efficiency shows a noticeable increase precisely at the source (as indicated by the black curve).
\end{itemize}

\subsection{Oscillations around Charged  black holes in scalar-tensor Gauss-Bonnet gravity}

In discussing particle motion in an accretion disk, three distinct frequencies play a pivotal role. These frequencies are the vertical frequency $\Omega_\theta$, which controls harmonic vertical motion, the radial frequency $\Omega_r$, which regulates harmonic radial motion, and the orbital frequency $\Omega_\phi$, which drives circular motion within the disk. In order to measure the vertical and radial motions, we employ the relationships defined as $\frac{1}{2}\left(\frac{dr}{dt}\right)^2=V^{(r)}_{eff}$ and $\frac{1}{2}\left(\frac{d\theta}{dt}\right)^2=V^{(\theta)}_{eff}$. Here, Eq. (\ref{9}) is employed, yielding $u^\theta=0$ and $u^r=0$ for motion in radial and vertical direction, respectively. Using $u^r=\frac{dr}{d\tau}=\frac{dr}{st}u^t$ and $u^\theta=\frac{d\theta}{d\tau}=\frac{d\theta}{st}u^t$, we obtain the following expressions
\begin{eqnarray}\label{26}
\frac{1}{2}\left(\frac{dr}{dt}\right)^2&=&-\frac{1}{2}\frac{B(r)A^2(r) (r)}{E^2}\left[1+\frac{E^2}{A(r)}+\frac{L^2}{r^2\sin^2 \theta}\right]=V^{(r)}_ {eff},\\
\frac{1}{2}\left(\frac{d\theta}{dt}\right)^2&=&-\frac{1}{2}\frac{A^2(r)}{r^2E^2}\left[1+\frac{E^2}{A(r)}+\frac{L^2}{r^2\sin^2 \theta}\right]=V^{(\theta)}_ {eff},
\end{eqnarray}
where $\Omega^2 (\theta)=-\frac{d^2}{d\theta^2}V^{(\theta)}_ {eff}$. Then from Eq. (\ref{26}),
we obtain
\begin{widetext}
\begin{eqnarray}
\Omega^{2}_{\theta}=&&\frac{A^2 (r)L^2}{r^4 E^2},\label{30}\\
\Omega^{2}_{r}=&&\frac{1}{2r^4 E^2}[(r^2+L^2)r^2f^2(r)-E^2r^4A(r)]B''(r)+2[(r^2+L^2)f(r)-E^2r^2]r^2B(r)A''(r) \\\nonumber
&&+2r^2B(r)A'^2(r)(r^2+L^2)-2r\Big[[(r^2+L^2)2A(r)+E^2r^2]r B'(r)+4L^2A(r)B^2(r)\Big]\\\nonumber
&&\times A'(r)-4L^2f^2(r)(-\frac{3}{2}B(r)+r B'(r))].\label{31}
\end{eqnarray}
\end{widetext}
\section{Accretion Regime around Charged  black holes in scalar-tensor Gauss-Bonnet gravity}

In order to derive some general findings concerning accretion, we use the energy-momentum tensor corresponding to a perfect fluid. This is given by
\begin{equation}
T^{\mu\nu} =(p+\rho)u^\mu u^\nu + pg^ {\mu \nu},\label{32}
\end{equation}
where $\rho$ and $p$ are the energy density and isotropic pressure, respectively. Also, the general expression for the four-velocity is given by
\begin{equation}
u^\mu = \frac{dx^\mu}{d\tau}=(u^t,u^r,0,0),\label{33}
\end{equation}
where $\tau$ is the appropriate time associated with the geodesic movement of a particle. In the case of steady-state and spherically symmetric flow, the normalization condition $u^\mu u_\mu = -1$ is employed. Therefore, the four-velocity of the fluid can be expressed as
\begin{equation}
u^t =\Big[\frac{B(r)+(u^r)^2}{A(r)^2B(r)^2}\Big]^{1/2}.\label{34}
\end{equation}
 Inward flow, represented by $u^r<0$, causes accretion and corresponds to a negative velocity of the fluid. In contrast, in outflow with $u^r>0$, the fluid velocity is positive. Hence, when analyzing accretion, we are focusing on the conservation laws of energy and momentum, which can be expressed as
\begin{equation}
T^{\mu\nu}_{;\mu} =0 \quad\Rightarrow\quad \frac{1}{\sqrt{-g}}(\sqrt{-g}T^{\mu\nu})_{,\mu}+T^{\alpha\mu}\Gamma^{\nu}_{\alpha\mu}=0,\label{35}
\end{equation}
where the covariant derivative is represented by a semicolon. Further, the square root of the determinant of the metric tensor is denoted by $\sqrt{-g}$. Moreover, considering the second type of Christoffel symbol, we are able to derive the following expression
\begin{equation}
r^2u^r(\rho+p)\sqrt{B(r)+(u^r)^2}\frac{A(r)}{B(r)}=D_0,\label{36}
\end{equation}
where the term $D_0$ is an arbitrary constant. Applying the relationship between conservation law and four-velocity using the formula $u_{\mu}T^{\mu\nu}_{;\nu}=0$, we can obtain
\begin{equation}
(p+\rho)_{;\nu}u_{\mu}u^{\mu}u^{\nu}+(p+\rho)u^{\mu}_{;\nu}u_{\mu}u^{\nu}+(p+\rho)u_{\mu}u^{\mu}u^{\nu}_{;\nu}+p_{,\nu}g^{\mu\nu}u_{\mu}+p u_{\mu}g^{\mu\nu}_{;\nu}=0.\label{37}
\end{equation}
By supposing that $ u^\mu u_\mu = -1$ and $g^{\mu\nu}_{;\nu}=0$, the above relation can be simplified into
\begin{equation}
(\rho+p)u^{\nu}_{;\nu}+u^{\nu}\rho_{\nu}=0.\label{38}
\end{equation}
If we consider only the non-vanishing elements, we are left with
\begin{equation}
\frac{\rho'}{p+\rho}+\frac{1}{2}\frac{A'(r)}{A(r)}-\frac{1}{2}\frac{B'(r)}{B(r)}+\frac{u'}{u}+\frac{2}{r}=0.\label{39}
\end{equation}
By integrating, it gives
\begin{equation}
r^2u^{r}\sqrt{\frac{A(r)}{B(r)}}\exp\int\frac{d\rho}{p+\rho}=-D_{1},\label{40}
\end{equation}
where $D_{1}$ stands for the integration constant. Furthermore, considering the minus sign on the right-hand side, which indicates $u^r<0$, we arrive at the following expression
\begin{equation}
(p+\rho)\sqrt{A(r)\left[\frac{(u^r)^2}{B(r)}+1\right]}\exp\left(-\int\frac{d\rho}{p+\rho}\right)=D_{2}.\label{41}
\end{equation}
Here, the integration constant is denoted by $D_{2}$. This arrangement leads to the following expression
\begin{equation}
(\rho u^\mu)_{;\mu}\equiv\frac{1}{\sqrt{-g}}(\sqrt{-g}\rho u^\mu)_{,\mu}=0.\label{42}
\end{equation}
Alternatively, it can be expressed as
\begin{equation}
\frac{1}{\sqrt{-g}}(\sqrt{-g}\rho u^\mu)_{,r}+\frac{1}{\sqrt{-g}}(\sqrt{-g}\rho u^\mu)_{,\theta}=0.\label{43}
\end{equation}
The law of mass conservation can thus be expressed as
\begin{equation}
r^2\rho u^r\sqrt{\frac{A(r)}{B(r)}}=D_{3},\label{44}
\end{equation}
where a constant of integration appears again and is denoted by $D_{3}$.

\subsection{Dynamical parameters for Charged  black holes in scalar-tensor Gauss–Bonnet gravity}

Moving on, let us consider an isothermal fluid with the equation of state (EoS) $p=k\rho$, where $k$ denotes the state parameter. In the case of isothermal fluids, the temperature remains constant throughout the flow. Moreover, for such fluids, the sound speed remains constant, as the pressure is directly proportional to the density: $p\propto\rho$ throughout the accretion. By considering Eqs. (\ref{40}), (\ref{41}) and (\ref{44}), we can obtain the following expression
\begin{equation}
D_{4}=\Bigg[\frac{p+\rho}{\rho}\Bigg]\sqrt{A(r)\left[\frac{(u^r)^2}{B(r)}+1\right]}\exp\left(-\int\frac{d\rho}{p+\rho}\right),\label{45}
\end{equation}
with $D_4$ being an integration constant. By substituting the EoS $p=k\rho$ into the above expression, we obtain the general form of the radial velocity as
\begin{equation}
u=\Bigg[\frac{1}{k+1}\Bigg]\sqrt{B(r)\Big[\frac{(D_{4})^2}{A(r)}-(k+1)^2\Big]}.\label{46}
\end{equation}
Therefore, the radial velocity of AdS BH is given by
\begin{widetext}
\begin{eqnarray}\label{47}
u&=&\Bigg[\frac{1}{k+1}\Bigg]\sqrt{\Big(\Lambda r^2+1-\frac{2M}{r}+\frac{c^{5/2}_1}{r^5}\Big)\Big[\frac{(D_{4})^2}{\Big(\Lambda r^2+1-\frac{2M}{r}+\frac{2Mc_1}{r^3}-\frac{2Mc^2_1}{r^5}\Big)}-(k+1)^2\Big]}.
\end{eqnarray}
\end{widetext}
By using Eq. (\ref{44}), we can express the general form of the fluid density as follows
\begin{equation}\label{49}
\rho=\frac{D_{3}}{r^2}\frac{(k+1)}{\sqrt{{\frac{(D_{4})^2}{A(r)}-(k+1)^2}}}.
\end{equation}
Hence, the energy density of the strong and weak fields is given by
\begin{eqnarray}\label{50}
\rho&=&\frac{D_{3}}{r^2}\frac{(k+1)}{\sqrt{{\frac{(D_{4})^2}{\Big(\Lambda r^2+1-\frac{2M}{r}+\frac{2Mc_1}{r^3}-\frac{2Mc^2_1}{r^5}\Big)}-(k+1)^2}}}.
\end{eqnarray}

\begin{figure*}
\centering \epsfig{file=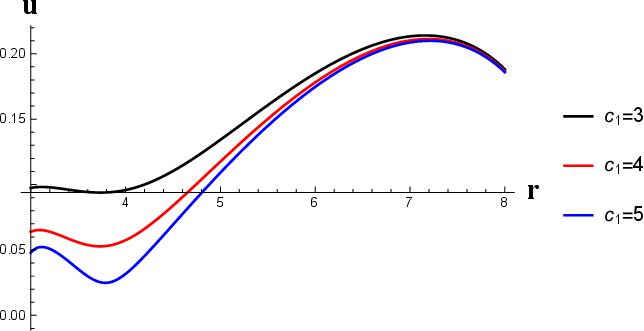, width=.40\linewidth,
height=2.0in}~~~~~~~~~~~~\epsfig{file=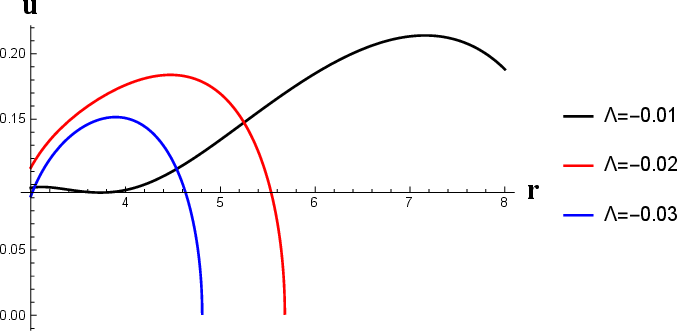, width=.40\linewidth,
height=2.0in}\caption{\label{F9} The graph illustrates the radial velocity of the CBH according to $r$ for different values of $c_1$ and $\Lambda$.}
\end{figure*}
The fluid velocity associated with the given parameters is shown in Fig. (\ref{F9}).  In the left graph, we can see that for the initial value of $c_1$, the velocity of the fluid increases with decreasing radius towards the source. On the other hand, for other values of $c_1$, the velocity decreases with increasing radius. In the right graph, it is observed that the velocity of the fluid is decreasing around the source and increasing away from the source. This behavior is related to the parameter $\Lambda$.

\begin{figure*}
\centering \epsfig{file=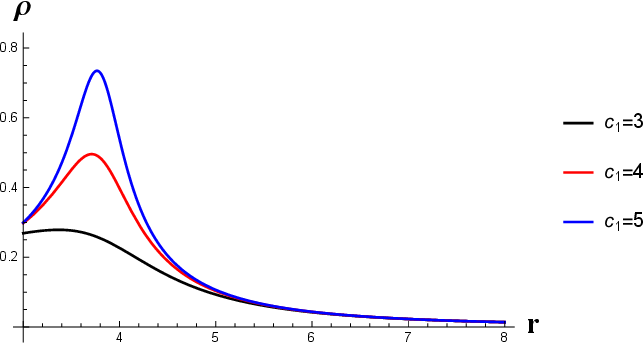, width=.40\linewidth,
height=2.0in}~~~~~~~~~~~~\epsfig{file=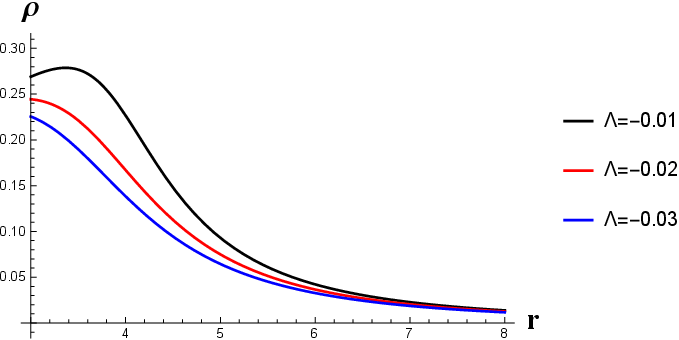, width=.40\linewidth,
height=2.0in}\caption{\label{F10} The graph displays the energy density of the CBH according to $r$ for different values of $c_1$ and $\Lambda$.}
\end{figure*}

The behavior of the energy density is described in Fig. (\ref{F10}), from which one can identify. some physical behavior. In the graph on the left, it is clearly seen that the fluid density increases with increasing parameter $c_1$, and, conversely, it decreases when $c_1$ decreases. In addition, all solution curves tend to converge near the source mass, pointing to a higher density in proximity to the source. In the right graph, we notice that increasing the parameter $\Lambda$ results in a decrease in fluid density, while decreasing $\Lambda$ results in an increase in density. In addition, all solution curves tend to converge around the source mass, reflecting a higher density closer to the source, whatever the specific value of $\Lambda$ is chosen.

\begin{figure*}
\centering \epsfig{file=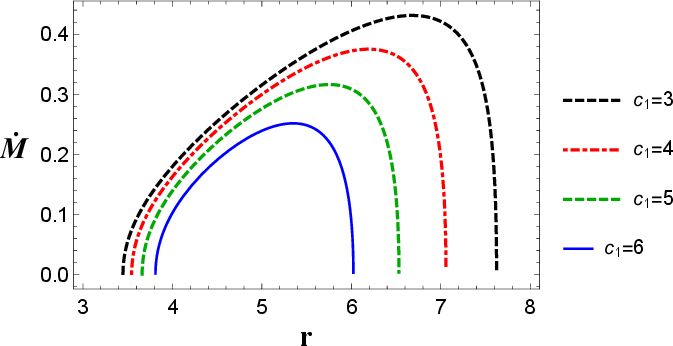, width=.40\linewidth,
height=2.0in}~~~~~~~~~~~~\epsfig{file=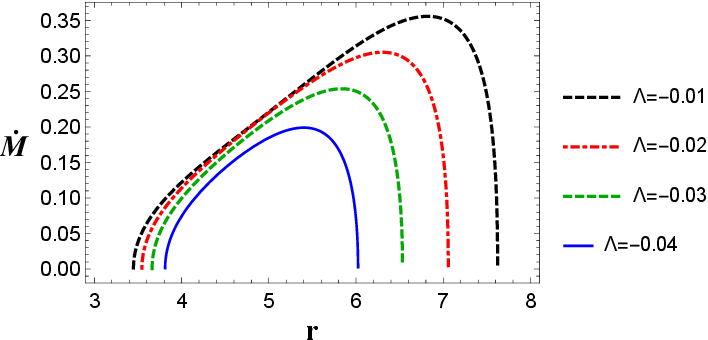, width=.40\linewidth,
height=2.0in}\caption{\label{F11} The graph shows the CBH mass accretion rate according to $r$ for different values of $c_1$ and $\Lambda$.}
\end{figure*}

The mass accretion rate for the CBH is examined in Fig. (\ref{F11}), and some key points can be observed.  In the left graph, it is clear that the maximum accretion rate of the CBH arises at the maximum radius when the initial value of $c_1$ is taken into account. For other values of $c_1$, on the other hand, the accretion rate decreases as the radius is reduced. In addition, all the solution curves in the plot originate from the event horizon and terminate at it, suggesting different radii along the accretion process. In the right graph, we see a profile similar to that in the left graph, but corresponding to the parameter $\Lambda$.  

\subsection{Mass expansion around Charged  black holes in scalar-tensor Gauss–Bonnet gravity}

The mass accretion rate ($\dot{M}$) of any BH depends on the area times the flux at its boundary. This amount reflects the rate at which the mass is accreted onto the BH per unit of time. The value of $\dot{M}$ depends on the nature of the accreting fluid and the metric parameters. Mathematically, it can be expressed as
\begin{equation}
\dot{M}=-4\pi r^2 u^r(\rho+p)\sqrt{A(r)+(u^r)^2}\equiv-4\pi D_0,\label{52}
\end{equation}
where $D_0=-D_1N_2$ and $D_2=(p_\infty+\rho_\infty)\sqrt{A(r_\infty)}$. Substituting these values into the above equation gives the following.
\begin{equation}
\dot{M}=4\pi D_{1}[\rho_{\infty}+p{_\infty}]\sqrt{A(r_{\infty})}M^2.\label{53}
\end{equation}
To investigate the rate of mass accretion, we consider a boundary at infinity, designated $r=r_{\infty}$. This boundary corresponds to the point where the massive particles fall into the BH. At this boundary, we have the energy density $\rho_{\infty}$ and the pressure $p_{\infty}$. Based on these considerations, the mass accretion rate formula can be stated in the following form 
\begin{equation}
\frac{dM}{M^2}=  \mathcal{F}dt,\label{54}
\end{equation}
where $\mathcal{F}=4\pi N_{1}(\rho_{\infty}+p{_\infty})\sqrt{A(r_{\infty})}$. By integrating, we get the following:
\begin{equation}
M_t=\frac{M_i}{1-Ft M_i}\equiv\frac{M_i}{1-\frac{t}{t_{cr}}}.\label{55}
\end{equation}
Here, $M_i$ denotes the initial mass of the BH, $M_t$ stands for the mass of the BH with the critical accretion time, and $t_{cr}$ is the evolution of the critical accretion time defined as
\begin{equation}
t_{cr}=\left[4\pi D_{1}[\rho_{\infty}+p{_\infty}]\sqrt{A(r_{\infty})M_i}\right]^{-1}.\label{56a}
\end{equation}
Consequently, the general expression for the BH mass accretion rate takes the form 
\begin{equation}
\dot{M}=4\pi D_{1}[\rho+p]M^2.\label{56}
\end{equation}

\section{Circular equatorial geodesics around Charged  black holes in scalar-tensor Gauss-Bonnet gravity}

The explicit forms for the effective potential, specific energy, and specific angular momentum of a moving particle play a vital role in the physical analysis. These quantities reveal valuable information about the dynamics and behavior of the particle in a system. Based on Eqs. (\ref{13}), (\ref{14}) and (\ref{15}), we can extract the following expressions
\begin{eqnarray}\label{58}
V_{eff}&=&\Bigg[\Lambda r^2+1-\frac{2M}{r}+\frac{2Mc_1}{r^3}-\frac{2Mc^2_1}{r^5}\Bigg]\Bigg[1+\frac{L^2}{r^2}\Bigg],
\\\label{64}
E^2&=&\frac{\Big[r^4(-2M+r+r^3\Lambda)+2M(r^2-c_1)c_1\Big]^2}{r^9[-3M+r]+Mr^5[5r^2-7c_1]c_1},
\\\label{66}
L^2&=&\frac{Mr^6+r^9\Lambda-3Mr^4c_1+5Mr^2c^2_1}{r^4[-3M+r]+M[5r^2-7c_1]c_1}.
\end{eqnarray}

Circular orbits are analyzable by assuming conditions in which both the effective potential and its derivative with respect to $r$ is equal to zero, i.e. $V_{eff}=0$ and $dV_{eff}/dr=0$. The ISCO is therefore governed by these conditions. For circular orbits to be stable, the second derivative of $V_{eff}$ versus $r$, noted $d^2V_{eff}/dr^2$, should be less than or equal to zero, i.e. $d^2V_{eff}/dr^2 \leq0$. Once the second derivative is equal to zero, it coincides with the location of the $ISCO$. Generally, $r_{ISCO}$ can be derived from
\begin{equation}
r_{ISCO}=\frac{3A'(r_{ISCO})A(r_{ISCO})}{2A'(r_{ISCO})^2-A''(r_{ISCO})A(r_{ISCO})}.\label{60}
\end{equation}
This implies that
\begin{widetext}
\begin{eqnarray}\label{60x}
r_{ISCO}&=&r-\frac{3r\Bigg[r^4(-2M+r+r^3\Lambda)+2M(r^2-c_1)c_1\Bigg]\Bigg[Mr^4+r^7\Lambda-3Mr^2c_1+5Mc^2_1\Bigg]}
{r^9\Bigg[r^3\Lambda[-1+3r^2\Lambda]+2Mc_1[r^6[2M-6r-19r^3\Lambda]+2Mc_1[-9r^2+10c_1]]\Bigg]}.
\end{eqnarray}
\end{widetext}
The $ISCO$'s characteristic radius in the equatorial plane is designated by $r_{ISCO}$ and denotes the specific radius that complies with the conditions discussed above. The $ISCO$ is playing a major role in understanding the accretion process surrounding CBHs. Nevertheless, obtaining an explicit analytical solution for the characteristic radius $r_{ISCO}$ related to CBHs is challenging due to its complexity. Consequently, numerical methods, such as employing Wolfram Mathematica, are being used to investigate and derive numerical values for the $ISCO$ as presented in Table \ref{Table1}. 
\begin{table}[ht]
\caption{The numerical solution of $r_{ISCO}$, $L_{ISCO}$ and $E_{ISCO}$ for a test particle moving in a static spherically symmetric CBH associated with different values of $c_1$ and fixed values of the parameters $M=1$ and $\Lambda=-1$.}\label{Table1}
\begin{center}
\begin{tabular}{|c|c|c|c|c|c|}
\hline \textbf{$c_1$} &\textbf{ $r_{ISCO}$}& \textbf{$L_{ISCO}$} & \textbf{$E_{ISCO}$}
\\\hline  $1$ &2.9417 & $383.433$ & $471.575$
\\\hline  $2$ &2.5398 & $67.3853$ & $224.308$
\\\hline  $3$ &2.7439 & $117.004$ & $220.332$
\\\hline
\end{tabular}
\end{center}
\end{table}

\subsection{Epicyclic frequencies around Charged  black holes in scalar-tensor Gauss–Bonnet gravity}

The particles moving in a circular orbit undergo small oscillations in the vertical and radial directions, which are considered to be special effects resulting from disturbances. Consequently, the frequencies corresponding to these oscillations can be given by 
\begin{eqnarray}\label{72}
\Omega^2_{\theta}&=&\frac{M}{r^3}+\Lambda-\frac{3Mc_1}{r^5}+\frac{5Mc^2_1}{r^7},\\\label{73}
\Omega^2_{r}&=&\frac{5M}{r^3}-\Lambda r-\frac{3Mc_1}{r^5}+M[5r^2-7c_1]c_1.
\end{eqnarray}
\begin{figure*}
\centering \epsfig{file=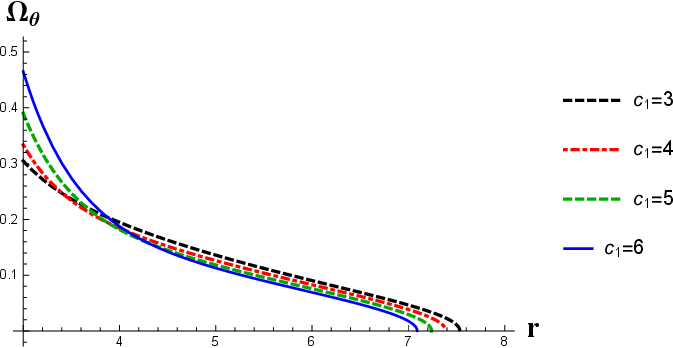, width=.40\linewidth,
height=2.0in}~~~~~~~~~~~~\epsfig{file=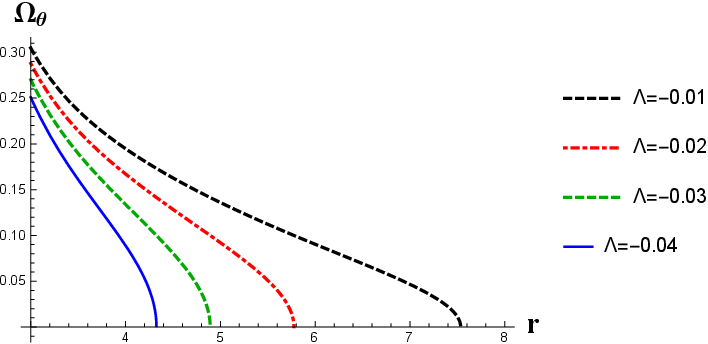, width=.40\linewidth,
height=2.0in}\\ \hspace{3.5cm}
\centering \epsfig{file=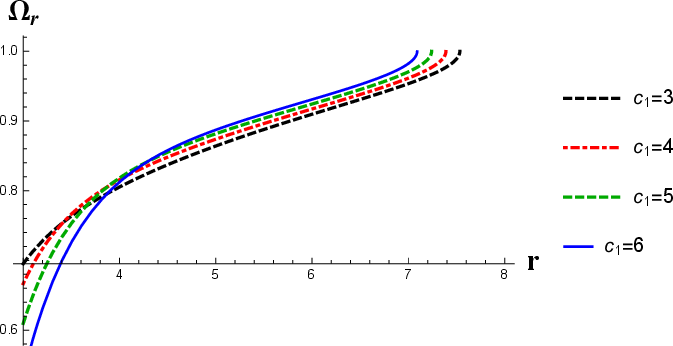, width=.40\linewidth,
height=2.0in}~~~~~~~~~~~~\epsfig{file=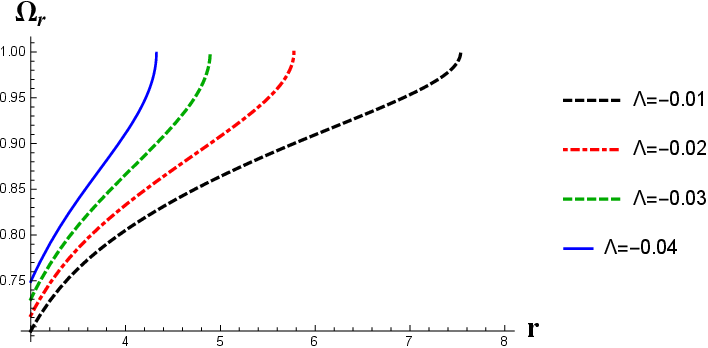, width=.40\linewidth,
height=2.0in}\\ \hspace{3.5cm}
\centering \epsfig{file=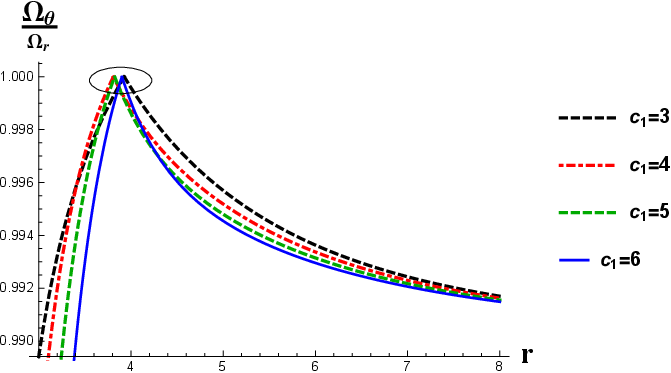, width=.40\linewidth,
height=2.0in}~~~~~~~~~~~~\epsfig{file=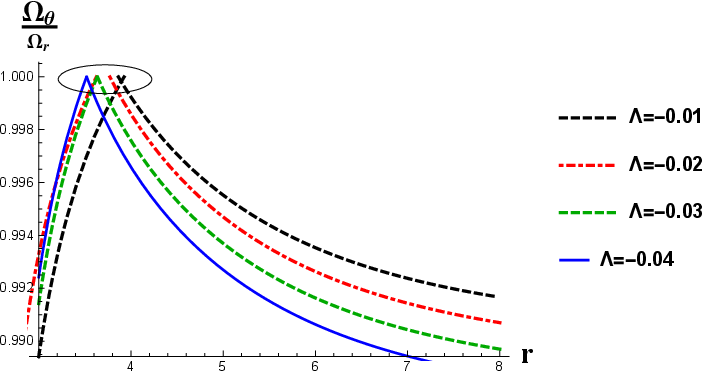, width=.40\linewidth,
height=2.0in}\caption{\label{F12} The graph illustrates the epicyclic frequencies of the CBH according to the radial coordinate $r$, considering different values of $c_1$ and $\Lambda$. The epicyclic frequencies show the characteristic oscillation rates associated with particle motion surrounding the CBH.}
\end{figure*}

Figure (\ref{F12}) shows the vertical and radial frequencies, along with their ratios, corresponding to the given parameters. The graph visually shows the relationship between these frequencies, providing useful information on their relative magnitudes and how they vary from each other. By increasing the parameter $c_1$, the vertical frequency increases and, in inversely, it decreases when $c_1$ is decreased. This trend can be seen in the upper left graph. In the upper right graph, with the progressively increasing parameter $\Lambda$, the vertical frequency becomes a decreasing function of the radial distance $r$. This relationship shows that with increasing $\Lambda$, the vertical frequency decreases for larger values of $r$. In the middle left graph, when the parameter $c_1$ is raised, the radial frequency is reduced. This reduction in radial frequency appears even for very small variations in radius. In the middle right graph, with increasing the parameter $\Lambda$, the radial frequency increases. This increase in frequency occurs without any change in radius.  In the lower two graph, we observe that the ratio $\frac{\Omega_{\theta}}{\Omega_{r}}$ stays constant, which implies that there is no change in energy with increasing or decreasing radius, depending on the two parameters $c_1$ and $\Lambda$.

\section{Dynamics of BHL Accretion in Scalar-Tensor Gauss-Bonnet Charged Black Hole Geometry}
\label{BHL_accretion} 

In addition to the theoretical analysis presented above, we perform numerical simulations to reveal the influence of modified gravity parameters on the BHL accretion mechanism and to examine how gravity affects the physical processes occurring in its vicinity. Modeling the BHL mechanism around a charged black hole is therefore essential to understand how the scalar–tensor Gauss–Bonnet corrections ($c_1$) and the cosmological parameter ($\Lambda$) modify the well-known shock cone structure that forms around a Schwarzschild black hole, as well as their impact on the density of the matter trapped inside the cone and its overall physical morphology. Comparing the outcomes obtained in modified gravity with those of the Schwarzschild case helps us to identify the imprints of the scalar–Gauss–Bonnet effects and the resulting gravitational field geometry on the accreting matter as it falls toward the black hole.

\begin{figure*}[!ht]
  \vspace{1cm}
  \centering
 \psfig{file=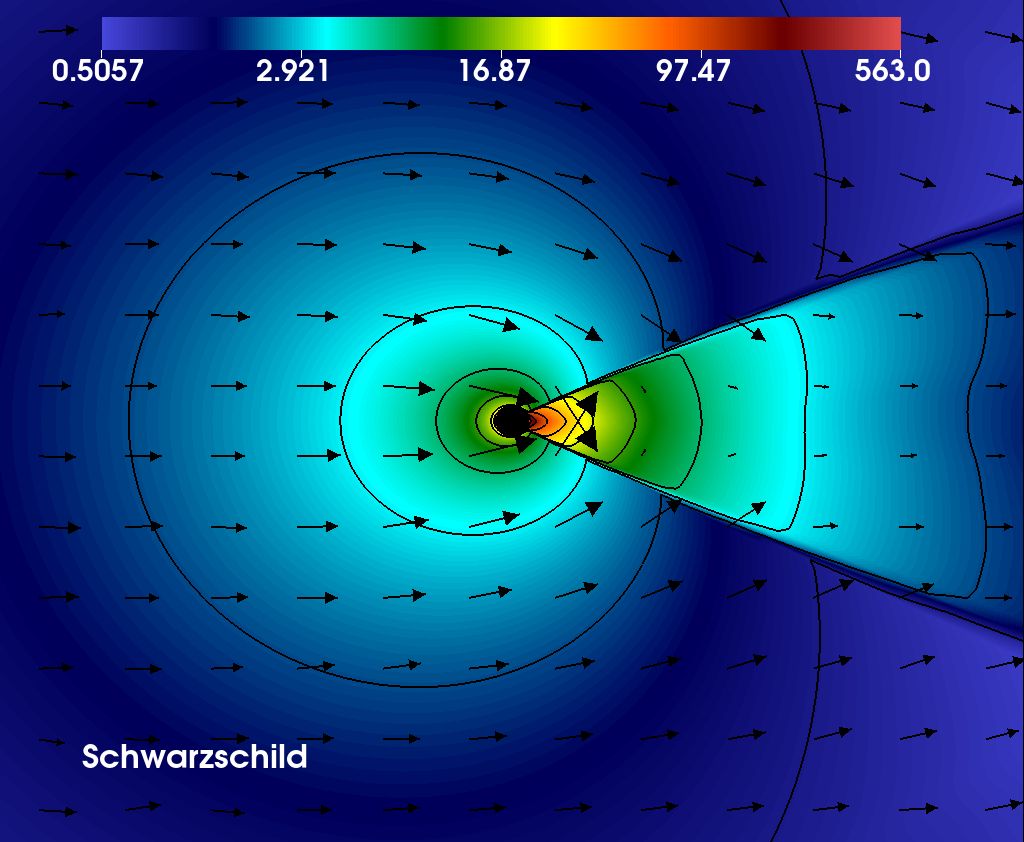,width=8.0cm,height=7cm}
 \psfig{file=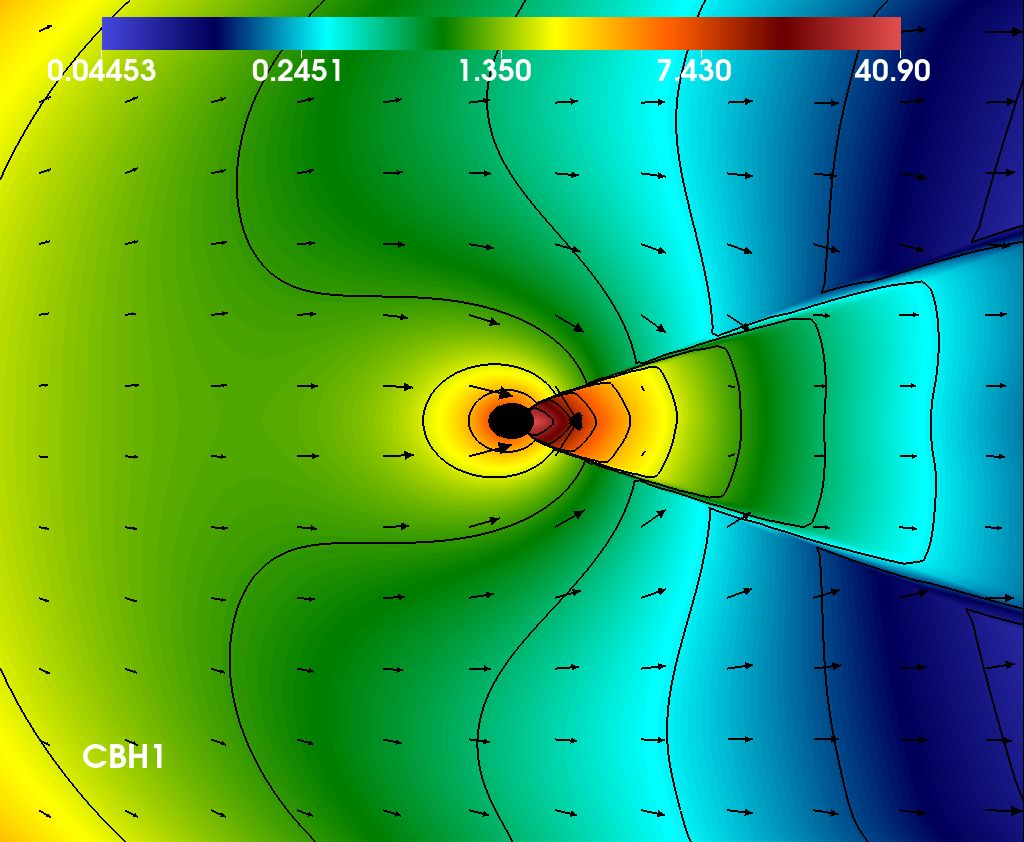,width=8.0cm,height=7cm}\\
 \psfig{file=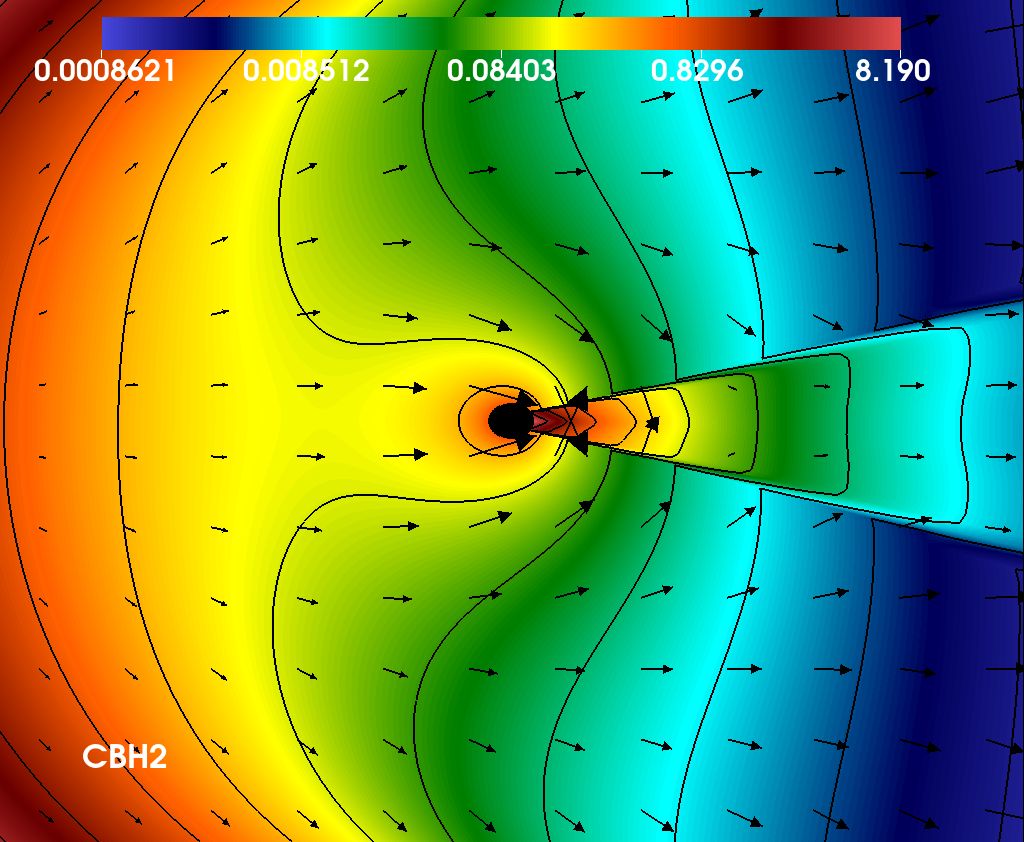,width=8.0cm,height=7cm}
 \psfig{file=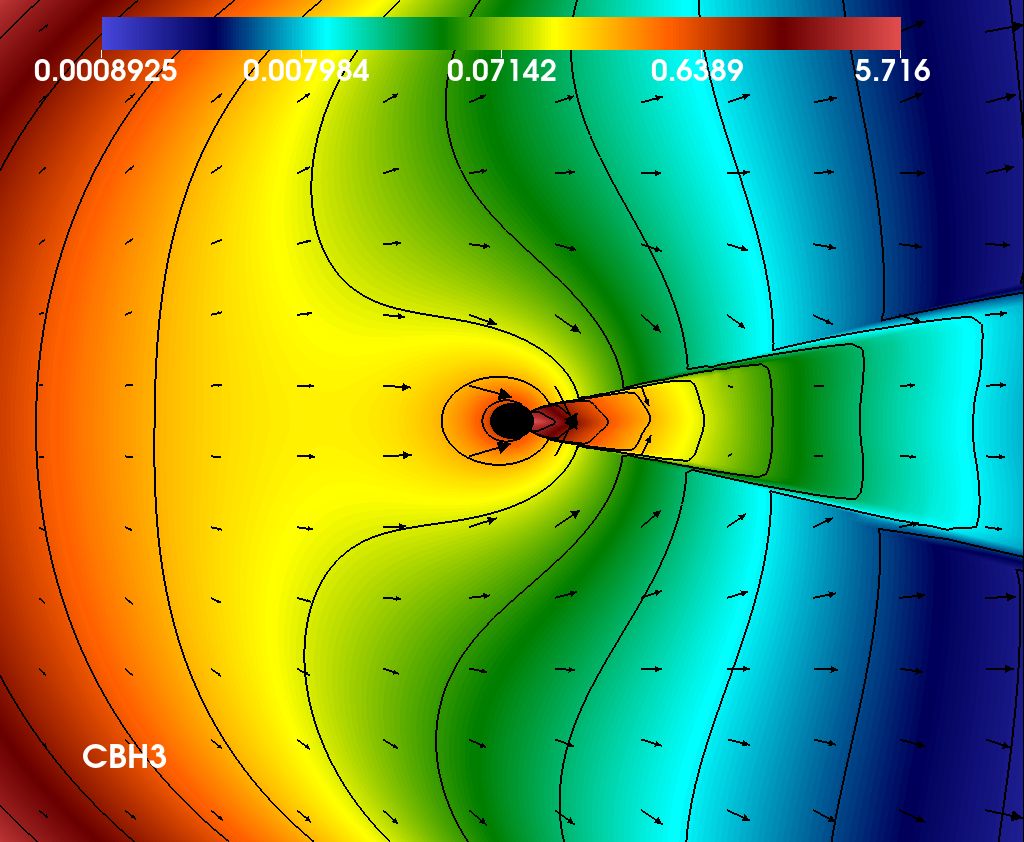,width=8.0cm,height=7cm}
  \caption{The figure shows the rest-mass density and velocity field distributions resulting from BHL accretion around different black hole geometries. Each panel illustrates the morphology of the shock cone formed around the black hole. The initial parameters corresponding to these panels are given in Table \ref{Inital_Con}. When compared with the Schwarzschild case, it is observed that the scalar–Gauss–Bonnet effects $c_1$ and cosmological constant $\Lambda$ have a significant influence on the formation and morphology of the shock cone.
}
  \vspace{1cm}
  \label{color_plots}
\end{figure*}

As in our previous studies \cite{Donmez2012MNRAS,KoyuncuMpla2014,Donmez2014MNRAS,Donmez2017MPLA,Donmez2024MPLA}, we numerically solve the GRHD equations to model the formation of the shock cone resulting from BHL accretion around black holes. The BHL accretion mechanism, as described in our previous publications \cite{DonmezJheap2024, Donmezpdu2024,DonmezEpjc2025,MustafaJcap2025}, is produced by injecting matter from the outer boundary toward the black hole, leading to the development of a downstream accretion flow. In this work, the spacetime into which the matter falls is defined using the initial conditions given in Table \ref{Inital_Con}. Fig. (\ref{color_plots}) presents the morphology of the shock cone that forms in the downstream region around the black hole under different parameter settings, using the initial values listed in Table \ref{Inital_Con}. The comparison with the Schwarzschild case clearly demonstrates how the dynamic structure of the shock cone changes with varying parameters.

\begin{table}[htbp]
\footnotesize
\caption{
}
 \label{Inital_Con}
\begin{center}
  \begin{tabular}{ccc}
    \hline
    \hline
 
     Model Name & \hspace{0.5cm} $\Lambda \left(1/M^{2}\right)$ \hspace{0.5cm} & \hspace{0.3cm} $c_1 \left(M^{2}\right)$   \\    
   \hline       
     Schwarzschild   &  $0$  &  $0$ \\
     CBH1            & $-5 \times 10^{-6}$ & $7$ \\
     CBH2            & $-1 \times 10^{-5}$ & $3$ \\     
     CBH3            & $-1 \times 10^{-5}$ & $7$ \\     
    \hline
    \hline
  \end{tabular}
\end{center}\label{tab1}
\end{table}

In the upper left panel of Fig. (\ref{color_plots}), the morphology of the shock cone formed in the Schwarzschild geometry is shown. As seen from the rest-mass density distribution, a sharply defined symmetric shock cone develops in the downstream region, opposite to the direction of the inflowing matter in the BHL mechanism. The formed shock cone is connected to the inner boundary of the computational domain at $r = 3M$. As matter approaches the event horizon, the gravitational force becomes stronger, causing the material trapped within the cone to accrete onto the black hole, leading to a region of high density near the horizon. The flow lines are nearly radial in the upstream region and converge strongly behind the black hole, forming a narrow and well-collimated shock cone. 

In the upper right panel of Fig. (\ref{color_plots}), the morphology of the shock cone for the CBH1 model is illustrated. In this case, where a mildly negative cosmological constant and a moderate Gauss–Bonnet coupling parameter are employed, it is evident that the modification of the gravitational potential significantly alters the structure of the shock cone. Compared with the Schwarzschild case, the opening angle of the shock cone in the CBH1 model slightly increases, and its boundaries become less sharply defined. The contrast between the density inside the cone and that of the ambient medium decreases, reflecting that the AdS-like cosmological background ($\Lambda < 0$) weakens gravitational confinement on large scales. Meanwhile, the Gauss–Bonnet coupling ($c_1 = 7$) affects the curvature of spacetime near the horizon. Compared with the Schwarzschild case, the velocity vector plots appear to be more curved, and the accretion streamlines are deflected earlier, indicating a weaker gravitational focusing and the presence of a broader wake. In general, compared to the Schwarzschild model, the CBH1 configuration demonstrates that the Gauss–Bonnet correction smooths the density gradient and partially diffuses the structure of the shock cone. 

In the CBH2 model, where the cosmological curvature parameter has a stronger influence on spacetime and the Gauss–Bonnet coupling is smaller ($c_1 = 3$), the corresponding shock cone morphology is shown in the lower left panel of Fig. (\ref{color_plots}). It is clearly observed that, compared to the Schwarzschild and CBH1 cases, the CBH2 shock cone exhibits a smaller opening angle (narrower shape), while its density distribution extends further away from the black hole. As shown in the two lower panels of Fig. (\ref{color_plots}), in the CBH2 and CBH3 models, the amount of matter falling toward the black hole and, consequently, the density of the material trapped inside the shock cone are significantly lower compared to the Schwarzschild case and slightly lower than in the CBH1 model. This indicates that the gravitational focusing around the black hole is not sufficiently strong, and therefore, the inflowing material is not effectively pulled or accreted toward the black hole. In fact, when examining the CBH1–CBH3 models, it can be clearly seen that near the outer boundary on the left side, where the matter initially flows inward through the BHL mechanism, the velocity vectors start to point outward, showing that the matter is being deflected away rather than falling in. In contrast, in the Schwarzschild case, the velocity field is directed inward, indicating efficient accretion. This explains why a smaller amount of material accumulates near the black hole in the CBH models. 

Consequently, a large portion of the matter that begins to fall toward the black hole through the BHL mechanism is redirected outward and is not captured by the black hole. This behavior can be attributed to the combined effect of the stronger AdS curvature (more negative $\Lambda$) and the modified Gauss–Bonnet coupling. The deeper AdS background alters the global pressure balance, while the curvature correction reduces the local gravitational attraction near the horizon. Together, these effects lead to weaker compression and insufficient mass accumulation around the black hole. As a result, both CBH2 and CBH3 display reduced accretion efficiency and lower central density compared to the other configurations.

\begin{figure*}[!ht]
  \vspace{1cm}
  \centering
 \psfig{file=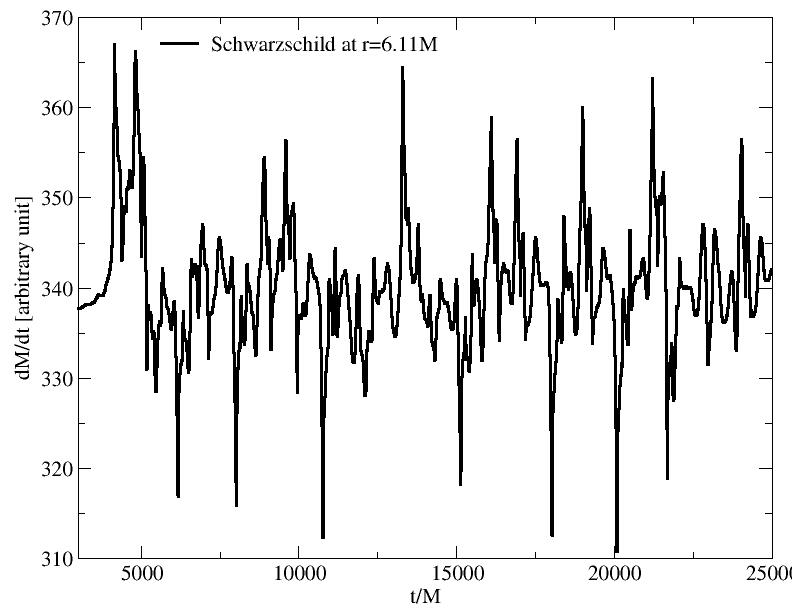,width=8.0cm,height=7cm}
 \psfig{file=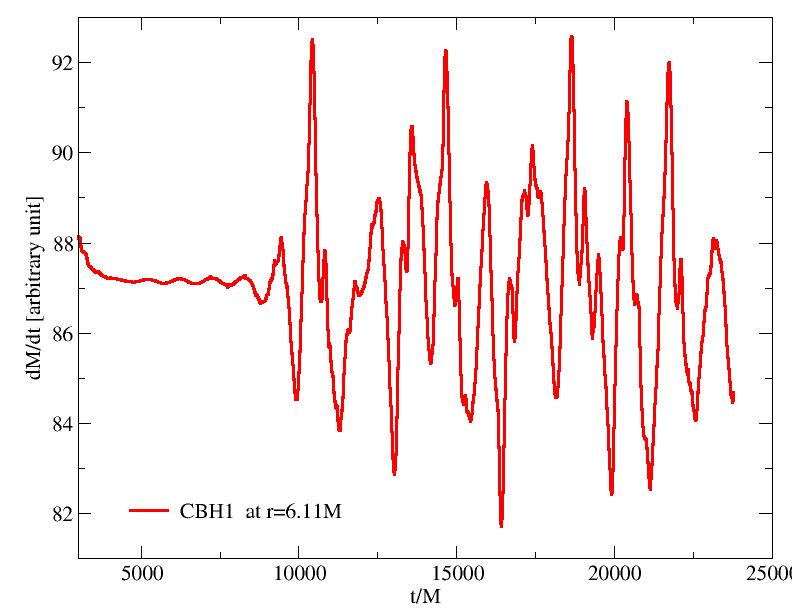,width=8.0cm,height=7cm}\\
 \psfig{file=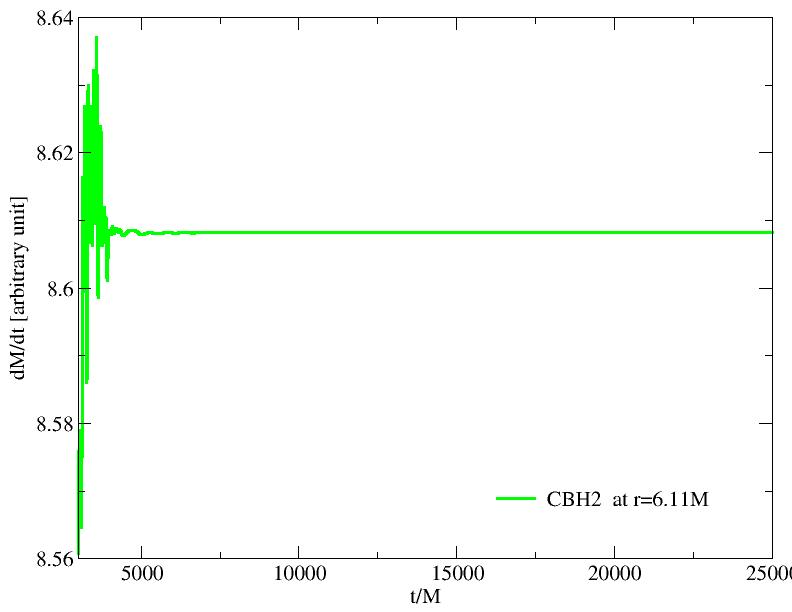,width=8.0cm,height=7cm}
 \psfig{file=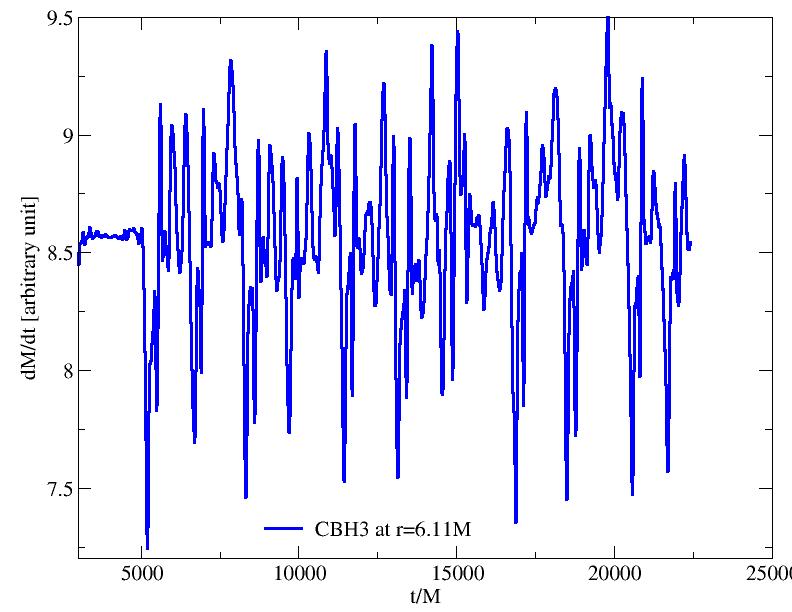,width=8.0cm,height=7cm}
  \caption{The time evolution of the mass accretion rate at $r = 6.11M$ is presented for the models given in Table \ref{Inital_Con}. It is observed that the amount of matter accreting around the black hole decreases significantly when compared with the Schwarzschild case, and consequently, the overall mass accretion rate also shows a strong reduction from model to model. Moreover, the oscillatory behavior of the accretion rate is found to be governed by the Gauss–Bonnet coupling and the cosmological parameters. Changes in these parameters clearly affect the strength, quasi-periodicity, and even the amplitude of the oscillations, demonstrating that they play a crucial role in regulating the temporal variability of the accretion flow.}
  \vspace{1cm}
  \label{mass_acc}
\end{figure*}

Fig. (\ref{mass_acc}) shows the time evolution of the mass accretion rate  $dM/dt$ at $r = 6.11M$ for different black hole models (Schwarzschild, CBH1, CBH2, and CBH3) given in Table \ref{Inital_Con}. In the upper left panel, corresponding to the Schwarzschild case, chaotic oscillations are observed after the shock cone reaches a quasi-steady state. This behavior indicates that in the strong gravitational region near the ISCO orbit, the matter trapped inside the shock cone exhibits strong oscillations. Since the coupling parameter is $c_1 = 0$ in the Schwarzschild geometry, the classical general relativistic potential allows the inflowing matter to fall directly toward the black hole, leading to significant variability in the accretion rate.  In the upper right panel (CBH1 model), a small coupling parameter $c_1$ and the presence of the cosmological parameter $\Lambda$ modify the gravitational potential, weakening the focusing effect.  Initially, no noticeable oscillations are observed, but later the system develops mild quasi-periodic variations. Compared with the Schwarzschild case, the oscillations persist but become smoother and less chaotic, showing transient peaks associated with shock oscillations. In the bottom left panel, corresponding to the CBH2 model, $c_1$ is smaller, but $\Lambda$ takes a larger negative value. Due to this stronger cosmological contribution, the accretion rate, which initially shows chaotic oscillations, rapidly stabilizes after the formation of the shock cone. After this short transient phase, it maintains steady laminar behavior throughout the entire simulation. This behavior shows that the modified gravitational potential significantly reduces the infall velocity and suppresses shock instabilities, producing a stable flow near the compact object.  Finally, in the bottom right panel (CBH3 model), where $c_1$ is larger than in CBH2 but $\Lambda$ remains the same, oscillations reappear after the initial transient region. However, since the amount of matter accreted toward the black hole is smaller, the oscillation amplitude is much lower compared to the Schwarzschild case.  Although these amplitudes are small, the appearance or disappearance of such oscillations is entirely the result of the coupling correction terms and the influence of the cosmological parameter. In conclusion, as both $c_1$ and $\Lambda$ increase, gravitational focusing weakens, the flow becomes smoother, and the mass accretion rate transitions from a turbulent, shock-dominated regime (Schwarzschild and CBH1) to a more stable and regulated accretion pattern (CBH2 and CBH3).  This shows that the coupling parameters act as damping factors controlling the strength of gravitational focusing.  These effects influence not only the BHL accretion mechanism itself but also the QPOs that appear after the system reaches the quasi-steady state.

\begin{figure*}[!ht]
  \vspace{1cm}
  \centering
 \psfig{file=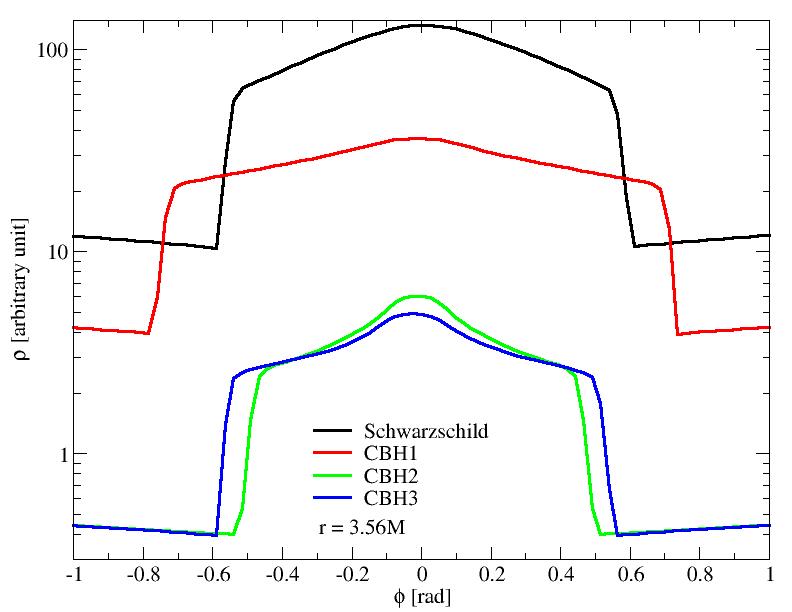,width=14.0cm,height=10cm}
  \caption{Rest-mass density variations along the azimuthal direction at $r = 3.56M$ for different black hole models. It is clearly observed that both the maximum value of the rest-mass density trapped inside the formed shock cone and the opening angle of the cone depend strongly on the Gauss–Bonnet coupling parameter $c_1$ and the negative cosmological constant $\Lambda$.}
  \vspace{1cm}
  \label{density_cut}
\end{figure*}

 Fig. (\ref{density_cut}) supports the behaviors observed in Figs.\ref{color_plots} and \ref{mass_acc}. In Fig. (\ref{density_cut}), the variation of the rest-mass density along the azimuthal direction at  $r = 3.56M$, close to the event horizon of the black hole, is presented. Thus, the shock locations formed in each model are clearly visible, allowing us to identify which model produces a wider cone and which model traps more matter inside the cone. As seen in Fig. (\ref{density_cut}), the Schwarzschild model exhibits the sharpest and highest rest-mass density, indicating strong gravitational focusing and efficient accretion of matter through the BHL accretion process. In contrast, the CBH1 model shows a lower and smoother density distribution, demonstrating that the scalar–Gauss–Bonnet coupling constant softens the gravitational potential, while the cosmological parameter reduces compression. As a result, the density of matter trapped inside the cone decreases, while the cone width increases significantly compared to the Schwarzschild case. In the CBH2 model, a greater reduction in density is observed because the more negative value of $\Lambda$ further weakens gravitational focusing, causing less matter to be accreted toward the black hole. With reduced accretion, the shock locations in the downstream region form closer to each other, significantly narrowing the cone opening angle. In the CBH3 model, the density inside the cone decreases even further, but the opening angle becomes slightly larger than in CBH2. The combined effect of a high $c_1$ and a strongly negative $\Lambda$ almost suppresses gravitational focusing and leads to a highly stable low-density flow. Consequently, Figs.\ref{color_plots}, \ref{mass_acc}, and \ref{density_cut} together demonstrate that increasing the negative value of the cosmological constant reduces gravitational focusing, leading to less matter falling toward the black hole and thereby decreasing the density of matter trapped within the cone. However, increasing the Gauss–Bonnet coupling term increases the cone opening angle. All these effects strongly influence the frequencies and observability of QPOs trapped inside the cone, suggesting that both $c_1$ and $\Lambda$ could potentially be tested through astrophysical observations.

In the case of the Gauss–Bonnet coupling parameter $c_1$ and the negative cosmological constant $\Lambda$, the numerical results obtained from modeling the structure of the shock cone around the black hole through the BHL accretion mechanism exhibit strong consistency with the analytical test-particle results. These results are entirely complementary to each other. In the test-particle model, as the Gauss–Bonnet coupling parameter $c_1$ increases or a larger negative cosmological constant $\Lambda$ is adopted, it is observed that gravitational focusing decreases, the specific energy reduces, the energy efficiency weakens, and the ISCO radius shifts toward smaller values. These variations cause matter around the black hole to experience a weaker gravitational attraction and form less compact circular orbits. The numerical BHL accretion results confirm all these analytical physical findings: as $c_1$ and the negative value of $\Lambda$ increase, the shock cone around the black hole becomes broader and less dense, the density of matter trapped inside the cone decreases significantly, the oscillations become smoother or vanish entirely, and the mass accretion rate drops. In both the numerical and theoretical results, the modified gravity parameters act as damping factors, leading to a smaller amount of matter falling toward the black hole and driving the system toward a more stable behavior. Therefore, the theoretical test-particle dynamics and the full hydrodynamic simulations are in excellent qualitative and quantitative agreement. Both demonstrate that scalar–tensor Gauss–Bonnet corrections weaken gravitational focusing, lower accretion efficiency, and produce more stable, low-density accretion flows compared to the Schwarzschild case.


\section{Conclusion}
The testing of scalar-tensor GB gravity through circular orbits and accretion disks is becoming an incredibly exciting area of research. This alternative theory of gravity, including a scalar field coupled to the GB term, opens up a novel paradigm for exploring gravitational phenomena. By using circular orbits and accretion disks, we have successfully shown their effectiveness as experimental tools for testing the scalar-tensor GB theory in regimes featuring strong gravitational fields and high curvatures. Through the utilization of these techniques, we have also derived constraints on the parameters characterizing this theory, improving our insight into its validity and range of applicability. Moreover, we have been exploring general solutions for the geodesic motion of particles in a static, spherically symmetric spacetime, emphasizing the importance of the conserved quantities of energy ($E$) and angular momentum ($L$) in establishing particle trajectories. Within the scenario of circular motion in an equatorial plane, assuming a constant radial component $r$ means that there is no change in radial motion or radial velocity.

In considering an isothermal fluid with an equation of state $p=k\rho$, where $k$ stands for the parameter of state, we find that the temperature remains constant throughout the accretion flux. The mass accretion rate ($\dot{M}$) for a black hole relies on the surface and the flux at its boundary, mirroring the rate of mass accretion per unit time. We also discovered that the precise value of $\dot{M}$ is governed by the characteristics of the accretion fluid and the metric parameters. By investigating the effective potential, specific energy, and specific angular momentum of moving particles, we have obtained precious insights into their dynamics and behavior within the system. Furthermore, we have established that particles in circular orbits display small oscillations in the vertical and radial directions, which are attributed to perturbations.

The numerical modeling of the shock cone formation through BHL accretion around the scalar–tensor Gauss–Bonnet charged black hole has demonstrated that the Gauss–Bonnet coupling constant ($c_{1}$) and the negative cosmological constant ($\Lambda$) have a significant influence on the strength of accretion, the morphology of the formed cone, and its overall stability. The increase in magnitude of both $c_{1}$ and $\Lambda$ leads to a weaker gravitational focusing near the black hole, a considerable change in the opening angle of the shock cone, a smoother density gradient, and a substantial reduction in the mass inflow toward the downstream region. In contrast, the shock cone formed around a Schwarzschild black hole remains well-defined, dense, and highly turbulent. However, in the modified gravity models, the structure becomes more diffused, the density of matter trapped inside the cone is strongly reduced, and the oscillations in the mass accretion rate are significantly weakened or completely disappear. Consequently, the parameters $c_{1}$ and $\Lambda$ drive the accretion flow from a chaotic, shock-dominated regime into a more laminar and steady configuration. These remarkable transitions in the accretion dynamics also lead to a noticeable suppression of the QPO modes trapped inside the shock cone and reduce their potential observability.

The results obtained from numerical simulations reveal that the scalar–Gauss–Bonnet correction terms act as a gravitational damping parameteres, regulating the dynamics of matter accreting toward the black hole by altering both the local curvature near the horizon and the large-scale pressure distribution governed by $\Lambda$. Consequently, the combined influence of a strong Gauss–Bonnet coupling and an AdS-like background leads to a noticeable decrease in the amount of mass accumulated near the black hole, a reduction in energy conversion efficiency, and a smoother temporal evolution of the accretion flow. These behaviors, consistently observed in both the theoretical test-particle analyses and the numerically modeled shock cone structures, demonstrate that the inclusion of higher curvature coupling terms systematically reduces the effective gravitational attraction and modifies the behavior of matter in strong field regimes. Finally, the results presented here indicate that future high resolution observations of accretion dynamics and physical processes around black holes can provide valuable constraints on the parameters of scalar tensor Gauss–Bonnet gravity and may even help identify its potential imprints in astrophysical systems.

In summary, the testing of scalar-tensor GB gravity through circular orbits and accretion disks demonstrates a promising path for improving our knowledge of gravitational physics. The constraints resulting from these experiments, together with the analysis of particle dynamics and accretion processes, are contributing to the ongoing exploration of this alternative theory and its implications for our grasp of the cosmos.

\section*{Acknowledgments}
AE thanks the National Research Foundation (NRF) of South Africa for the award of a postdoctoral fellowship.

\bibliographystyle{apsrev4-1}  
\bibliography{Accretion}

\end{document}